\documentclass[sigconf, nonacm]{acmart}
\renewcommand\footnotetextcopyrightpermission[1]{} %
\makeatletter %
\def\@formatdoi#1{} %
\makeatother %
\AtBeginDocument{%
  }

\usepackage{graphicx}
\usepackage{subfig}
\usepackage{xcolor}
\usepackage{tikz}

\definecolor{usedlink}{HTML}{E66101}
\definecolor{unusedlink}{HTML}{D0D0D0}
\definecolor{highlightlink}{HTML}{C51B7D}
\definecolor{highlightnode}{HTML}{F4A6C1}

\usepackage{booktabs}
\usepackage{multirow}
\usepackage{tabularx}
\usepackage{hyphenat}
\usepackage{enumitem}
\usepackage{mfirstuc}
\usepackage{natbib}
\usepackage{cleveref}
\usepackage{footnote}
\usepackage{svg}
\usepackage{afterpage}
\newcommand{\smalltitle}[1] {\vspace{6pt} \noindent \textbf{#1}}

\definecolor{revisecolor}{RGB}{0,150,120}

\newcommand{\fix}[1]{{\color{black}{#1}}}

\definecolor{newmagenta}{RGB}{170,0,120}
\newcommand{\new}[1]{{#1}}

\begin{document}

\title{Towards Scale-Up Networking: Challenges and Open Questions}


\author{Tingting Xu$^1$, Zixun Su$^1$, Xinyang Huang$^1$, Bowen Liu$^1$, Kun Tan$^2$, Dong Lin$^3$, Hao Wang$^3$, \\ Yinben Xia$^4$, Weifeng Zhang$^4$, Binzhang Fu$^5$, Shuihai Hu$^5$, Kai Chen$^1$}
\authornote{Corresponding author, kaichen@cse.ust.hk}
\affiliation{%
\institution{$^1$iSING Lab, Hong Kong University of Science and Technology, $^2$Unaffiliated, $^3$Huawei, $^4$Tencent, $^5$Alibaba Group}
  \country{}
}






\renewcommand{\shortauthors}{}

\begin{abstract}

The rapid growth of AI models and the widespread use of distributed training and inference have made inter-accelerator communication a critical system bottleneck. Scale-up fabrics address this bottleneck by combining high per-accelerator bandwidth with low and predictable latency over tightly coupled accelerator domains. However, the appropriate topology, protocol mechanisms, communication semantics, and system integration remain actively debated.

We analyze the critical-path communication generated by parallelism strategies and derive the resulting requirements for domain scale, per-accelerator bandwidth, and collective completion latency. We then use these requirements as a framework to examine representative proprietary and open scale-up fabrics, revealing how their physical interconnects, topologies, protocols, and semantics make different trade-offs among performance, scalability, compatibility, and implementation cost.
Moreover, based on this analysis and discussions with industry experts, we identify challenges and open problems in physical interconnects and switching, efficient communication, semantics, resource sharing and heterogeneous device interconnection. This paper provides a requirement-driven foundation for analyzing current scale-up fabrics and guiding the design of future systems.

\end{abstract}





\maketitle

\section{Introduction}
\label{sec:1}
Large language models (LLMs) impose demanding bandwidth and latency requirements on accelerator interconnects. Their growing memory and computation demands have made multi-accelerator execution standard practice for both training and inference ~\cite{cottier2025risingcoststrainingfrontier,Sevilla_2022,
kaplan2020scaling,zhao2023survey,rasley2020deepspeed,MetaScale,TPUv4}. Recent open models further reinforce this trend: DeepSeek-V4-Pro comprises 1.6 trillion total parameters, while Kimi K3 reaches 2.8 trillion parameters~\cite{deepseekv4,kimik3}. When such models are distributed through tensor, pipeline, or expert parallelism (TP, PP, EP), accelerators repeatedly exchange activations, partial tensor results, and expert inputs~\cite{huang2019gpipeefficienttraininggiant, narayanan2019pipedream, rasley2020deepspeed, lepikhin2020gshard, MixNet, popfetcher,taccl,  autoccl, jiang2024megascale, teslamotors2024ttpoe,deepseekv32024,nccl228docs}.

These communication patterns motivate three closely related requirements for accelerator interconnects. First, increasingly large models and parallel deployments require connectivity beyond a small host-local configuration. Second, the large messages exchanged during training and prefill require high sustained per-accelerator bandwidth. Third, the frequent and dependency-critical TP collectives and EP all-to-all operations require low and predictable communication latency, particularly during online inference, where communication delays directly affect time to first token (TTFT) and time per output token (TPOT)~\cite{deepseekv32024,lepikhin2020gshard,rasley2020deepspeed,DistServe}.
Scale, bandwidth, and latency are therefore coupled requirements rather than independent properties: the fabric must sustain the required communication performance at the scale of the workload's communication-critical groups. We quantify these scale, bandwidth, and latency requirements in Section~\ref{sec:2}. 
\begin{figure}[t]
    \centering
    \includegraphics[width=\linewidth]{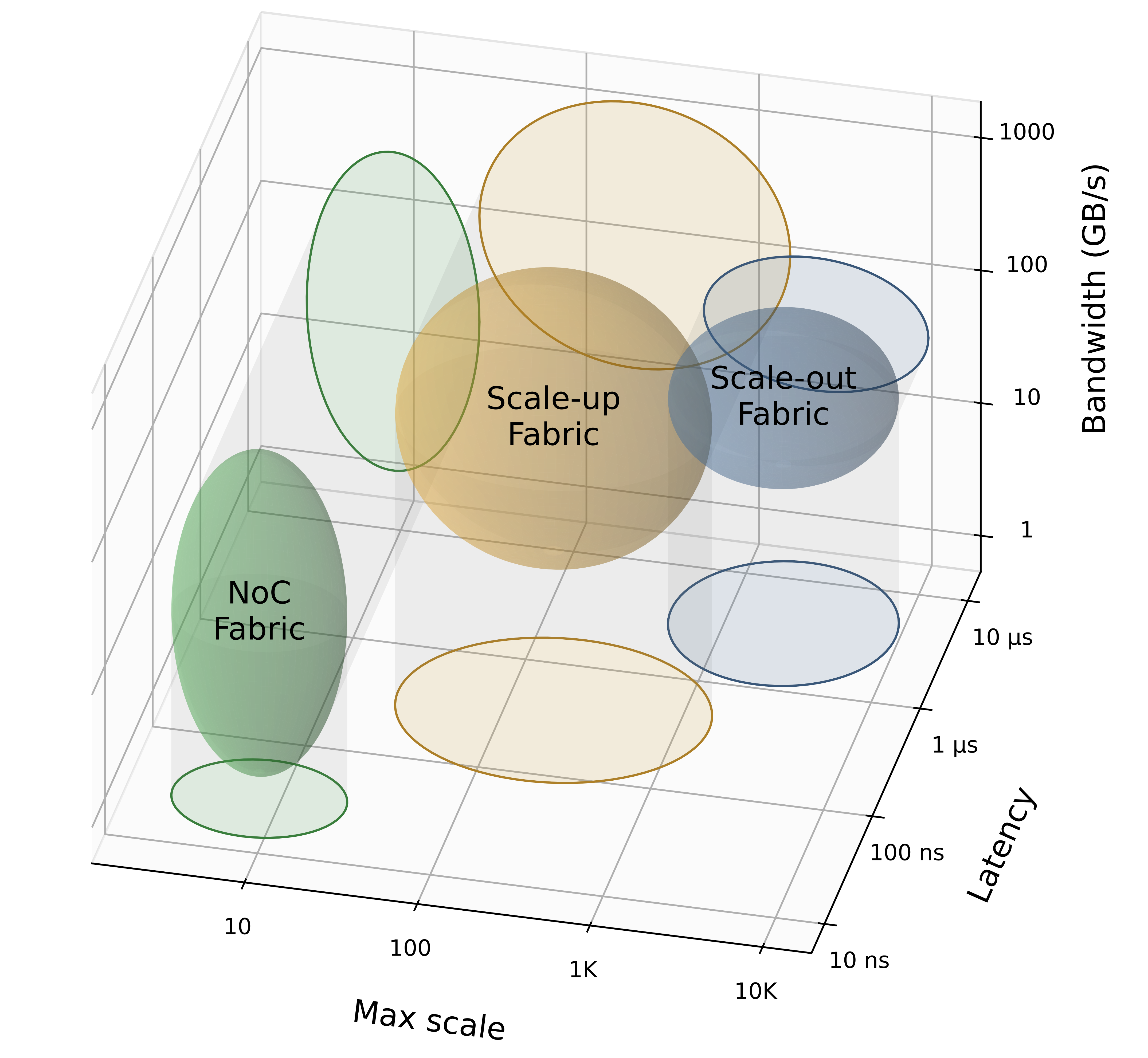}
    \caption{Positions of NoC/scale-in fabric, scale-out fabrics and scale-up fabrics in the bandwidth-latency-scale 3D map. 
    }
    \label{fig:map}
\end{figure}

Figure~\ref{fig:map} positions NoC (Network on Chip) or scale-in fabrics, scale-out networks, and scale-up fabrics along three dimensions: accelerator scale, per-accelerator bandwidth, and communication latency. Although all three classes can employ high-speed physical links, their different design objectives lead to substantially different effective bandwidth between accelerators. NoC and scale-in fabrics provide tightly integrated communication within a chip, package, board, or small compute complex. Their short physical paths and close integration enable high bandwidth and low latency, but endpoint degree, package I/O, board routing, and physical reach constrain their scale. Scale-out networks connect thousands or more endpoints, but NIC sharing, multi-stage forwarding, and network-wide contention limit their sustained per-accelerator bandwidth and latency predictability.

Modern AI workloads increasingly require high per-accelerator bandwidth, low and predictable communication latency, and connectivity beyond a single host, a combination that neither NoC fabrics nor scale-out networks can simultaneously provide. These emerging requirements have motivated scale-up fabrics, which aggregate multiple high-speed links per accelerator to provide hundreds of GB/s of bandwidth while retaining short communication paths and low collective completion latency~\cite{SUE,UALink,nvidiaBlackwell}. The scale of a scale-up fabric should not be defined by a fixed endpoint count. The required domain size depends on the workload, accelerator architecture, and parallelism mapping. Its effective scale is therefore the number of accelerators over which the fabric can sustain the required per-accelerator bandwidth and bounded communication latency.

\fix{
\smalltitle{Terminology and Scope.}
Terminology for scale-up networking is not yet fully standardized across academic and industrial sources. We therefore summarize the key terms used in this paper in Table~\ref{tab:terminology-def}. }

\begin{table*}[t]
  
  \centering
  \footnotesize
  \setlength{\tabcolsep}{5pt}
  \renewcommand{\arraystretch}{1.08}
  \caption{Terminology used throughout this paper.}
  \begin{tabularx}{\textwidth}{@{}>{\raggedright\arraybackslash}p{0.19\textwidth}
    >{\raggedright\arraybackslash}X@{}}
    \toprule
    \textbf{Term} & \textbf{Definition} \\
    \midrule
    Scale-up
      & One tightly-coupled accelerator system that preserves high bandwidth and low, predictable latency. Scale-up systems may extend across multiple nodes or racks; it is not restricted to adding resources within one server~\cite{SUE, UALink_spec}. A more detailed definition is laid out in Section~\ref{sec:2} \\ \cmidrule{2-2}
    Scale-out
      & Expansion by connecting additional, comparatively independent compute nodes or scale-up domains through a general network (usually through RoCE or Infiniband). Scale-out systems prioritize broad reach, large system size, and fault containment; Conventional data-center networks are a typical form of scale-out systems~\cite{alfares2008scalable, UEC2026Spec}. \\ \cmidrule{2-2}

    NoC / Scale-in
    & A tightly integrated communication fabric connecting processing, memory, or I/O components within a chip, package, board, or small compute complex. It generally provides short communication paths and high local bandwidth, while its scale is constrained by endpoint interfaces, packaging, board routing, and physical reach. \\ \cmidrule{2-2}
    
    Fabric
      & The communication substrate formed by physical links, switches, endpoint interfaces, protocols, and associated control or management mechanisms. A fabric provides connectivity and data-movement services; the attached compute and memory resources are not themselves part of the fabric. \\ \cmidrule{2-2}
    Domain
      & A bounded set of endpoints that are mutually reachable under a common scale-up fabric's addressing, routing, management schemes~\cite{SUE, UEC2026Spec}.  \\ \cmidrule{2-2}
    Endpoint
      & An addressable participant that acts as the source or destination of fabric transactions. Examples include accelerator ports, chiplets (IO dies) or memory nodes. A switch forwards transactions but is not an endpoint \cite{SUE, UECWhitepaper}. \\ \cmidrule{2-2}
    Accelerator
      & A vendor-neutral name for a workload-specialized compute engine. Typical examples include GPU, TPU, NPU, or other type of AI-related computing hardware \cite{SUE, UECWhitepaper}. In this paper, CPU is not considered an accelerator. \\ \cmidrule{2-2}
    Optical Interconnect
      & A link using optical signals as the primary media for conveying messages, including the associated optical engine, transceivers, fiber and other hardware.
        Electrical host interfaces may still be present at its endpoints \cite{kachris2012optical, miller2009optical}. \\ \cmidrule{2-2}
    Topology Reconfiguration
      & Runtime alteration of active paths or link capacities so that fabric connectivity better matches workload demand or avoids failed components \cite{TPUv4, TopoOpt}. \\ \cmidrule{2-2}

    
    Communication Collective &
    A communication operation defined over a group of participating endpoints, with specified data transformation, distribution, ordering, and completion semantics. Common examples include Broadcast, Reduce, AllReduce, AllGather, ReduceScatter, and All-to-All~\cite{mpiforum2023mpi41}.  \\ \cmidrule{2-2}

    In-network Computing & The execution of communication-related or application-defined operations on programmable fabric elements, such as switches or network interfaces. Examples include aggregation, reduction, data combination, filtering, and collective offload. In-network execution is constrained by switch computation, memory, state-management, and failure-handling resources~\cite{switchml}.\\ \cmidrule{2-2}
        
   Semantics
      & The externally observable contract exposed by a fabric to endpoints and software. It includes the supported operations, addressing model, initiation mechanism, ordering rules, completion guarantees, atomicity, memory-consistency behavior, and error or failure semantics. Examples include message passing, remote memory access, load/store operations, atomics, and collective operations. \\ \cmidrule{2-2}

    Memory Consistency &
    The set of rules that constrains the values a memory operation may observe and the apparent ordering of memory operations issued by one or more agents. A consistency model specifies which executions are permitted, including how program order, synchronization operations, and accesses to different memory locations constrain visibility \cite{sorin2020primer}.
     \\ \cmidrule{2-2}

    Cache Coherence &
    The property that accesses to an individual memory location observe a compatible ordering of writes across participating caches or agents. A coherence protocol propagates, invalidates, or otherwise coordinates cached copies so that stale and conflicting values are handled according to the architectural memory model. Coherence is primarily a per-location property and does not, by itself, define the ordering of operations to different locations~\cite{sorin2020primer,cxl2023spec31}. 
     \\ \cmidrule{2-2}
    Multi-tenancy / QoS Isolation
      & Multi-tenancy allows independent workloads to share a physical domain.
        Quality-of-service (QoS) isolation bounds cross-tenant interference so that one tenant cannot violate another's latency, bandwidth, or progress objective. \\ \cmidrule{2-2}
    Heterogeneous Domain
      & A scale-up domain assembled from endpoints that can differ in vendor, generation, compute capability, memory capacity and other metrics. \\
    \bottomrule
  \end{tabularx}
  
  \label{tab:terminology-def}
\end{table*}

In this paper, we examine eleven representative scale-up interconnect families spanning vendor-integrated systems, vendor-led technical disclosures, and consortium- or community-defined specifications. They include NVIDIA NVLink/NVSwitch~\cite{Nvidia_NVSwitch,NVLink}, AMD Infinity Fabric~\cite{InfinityFabric}, Google Inter-Chip Interconnect (ICI)~\cite{Google_TPU8,Jouppi2023TPUv4}, T-Head ICN~\cite{THeadICN,THeadICNSwitch}, Unified Bus 2.0~\cite{UBSpec,openeuler2025unifiedbus}, SUE/SUE-T~\cite{SUE_Whitepaper,SUE, SUE_T_Base_1_0}, UALink~\cite{UALink_spec,UALink,UALink_Webinar}, Tesla TTP/TTPoE~\cite{Quinnell2024ttpoe,teslamotors2024ttpoe}, ByteDance EthLink~\cite{EthLinkSpec}, Open Compute Project's Ethernet for Scale-Up Networking (ESUN)~\cite{OCP_ESUN_Launch,OCPESUN2026}, and High-Throughput Ethernet (ETH+)~\cite{ETHPlusSpec,ETHPlusConsortium}.
These entries cover different functional scopes and are not assumed to be equivalent full-stack fabrics.
Our analysis is organized around four complementary dimensions. First, \emph{topology and scale} considers how endpoints, links, switches, and fabric planes are organized to achieve the required bandwidth, latency, physical reach, and domain size. Second, \emph{communication and transport} covers physical- and data-link-layer mechanisms, packet formats, data transfer, flow control, reliability, routing, congestion management, and transport. Third, \emph{accelerator-visible semantics and execution models} examines supported operations, initiation paths, ordering, completion, visibility, and memory-consistency guarantees. Fourth, \emph{resource sharing and system integration} considers logical partitioning, access and performance isolation, management, and heterogeneous endpoint support. Together, these dimensions reveal how existing designs trade off performance, scalability, programmability, isolation, interoperability, implementation complexity, and cost.


Beyond analyzing existing solutions, we identify a set of challenges and open problems for future scale-up networks, informed by the design trade-offs exposed in this paper. These problems span four dimensions: improving the physical interconnect and switching fabric, enabling efficient communication within and across scale-up domains, supporting tightly coupled system semantics, and sharing and composing scale-up resources. Together, these problems highlight the difficulty of simultaneously providing low and predictable latency, high per-accelerator bandwidth, performance isolation, and tightly coupled communication as scale-up domains expand from small direct-connected systems to rack- and superpod-scale deployments.

This paper makes three contributions. First, it derives workload-dependent requirements for domain scale, per-accelerator bandwidth, and communication latency, and identifies the additional access- and performance-isolation requirements that arise when scale-up domains are shared. Second, it provides a structured analysis of eleven representative interconnect families across topology and scale, communication and transport, accelerator-visible semantics and execution models, and resource sharing and system integration. Third, it uses the resulting design trade-offs to formulate an application- and system-driven research agenda for future scale-up fabrics. We hope this framework provides a common foundation for comparing current designs and reasoning about emerging scale-up technologies.



\section{Scale-Up Fabrics}
\label{sec:2}

Existing scale-up fabrics differ substantially in topology, protocol architecture, communication semantics, and supported scale ~\cite{NVLink,TPUv4,SUE_Whitepaper,UALink_spec}. Rather than defining scale-up through a particular protocol or a fixed endpoint count, this section derives its requirements from modern AI workloads. We first examine LLM structure, scaling trends, and distributed execution. We then derive the global accelerator requirement from model capacity and computation, map DP, PP, TP, and EP onto scale-up domains, and analyze the communication requirements generated by these parallel dimensions. Finally, we synthesize these results into workload-derived requirements for domain scale, effective per-accelerator bandwidth, and collective completion time.

\subsection{Workload Trends and Parallel Execution}
\label{sec:workload_demand}
\begin{figure}
    \centering
    \subfloat[Data parallelism]{\includegraphics[width=0.45\linewidth]{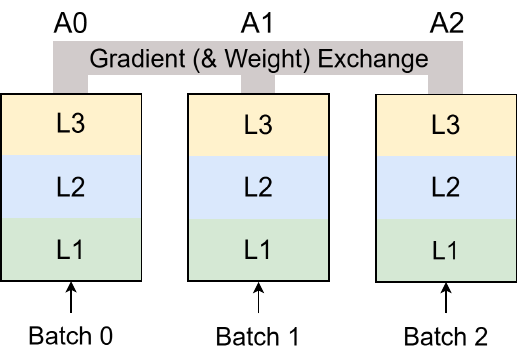}} \hfill
    \subfloat[Pipeline parallelism]{\includegraphics[width=0.45\linewidth]{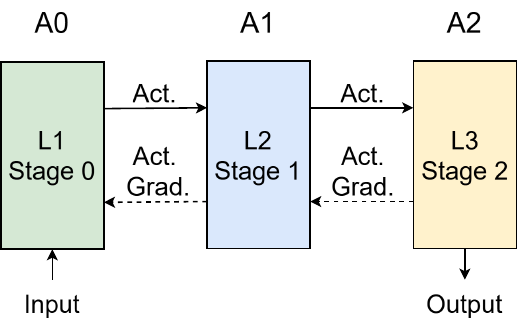}}
    \\
    \subfloat[Tensor parallelism]{%
    \includegraphics[
        width=0.45\linewidth,
        trim=0bp 2.74bp 0bp 25.5bp,
        clip
    ]{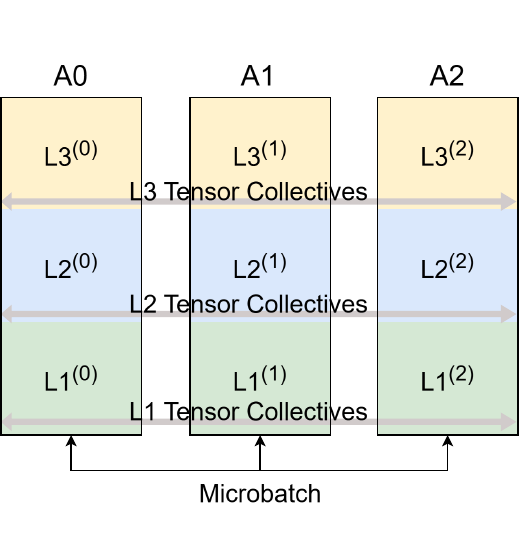}%
} \hfill
    \subfloat[Expert parallelism]{\includegraphics[width=0.45\linewidth]{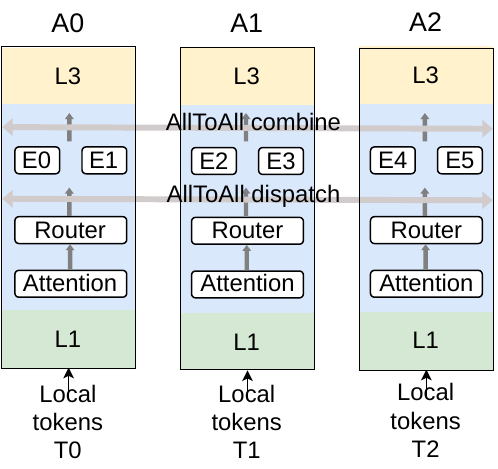}}
    \caption{Parallelism strategies. Grad. and Act. are abbreviations for gradients and activations.}
    \label{fig:parallelism}
\end{figure}
\smalltitle{Model structure and state.}
A modern LLM typically consists of an input embedding, a sequence of Transformer layers, and an output projection ~\cite{zhao2023survey,Llama3}. Each Transformer layer contains an attention module and a feed-forward network (FFN), together with normalization and residual connections. Let $L$ denote the number of layers, $H$ the hidden dimension, $H_{\mathrm{ff}}$ the FFN intermediate dimension, and $V$ the vocabulary size. For dense Transformers, the dominant parameter terms grow approximately with \(LH^2\) and \(LHH_{\mathrm{ff}}\). 

Mixture-of-experts (MoE) models replace the dense FFN in selected layers with multiple experts. If a layer contains $E$ experts and each token is routed to $k$ of them, the layer stores all $E$ experts but activates only $k$ experts per token. MoE models therefore distinguish \emph{total parameters}, which largely determine the model-state footprint, from \emph{active parameters}, which more directly determine per-token computation~\cite{lepikhin2020gshard,deepseekv32024}.

Parameters are only one component of accelerator memory consumption. Training additionally maintains gradients, optimizer states, activations, temporary workspace, and communication buffers \cite{ZeRO,rasley2020deepspeed}. Inference mainly stores weights, activations, runtime buffers, and a key-value (KV) cache. The KV cache grows with the number of layers, concurrent tokens, and context length, and can become a major capacity constraint during long-context or high-concurrency serving~\cite{DistServe}.

\smalltitle{Scaling trends.}
Empirical scaling laws show that model quality generally improves predictably as model size, training data, and computation increase ~\cite{kaplan2020scaling,Sevilla_2022}. Publicly disclosed models have progressed from hundreds-of-billions-parameter dense models to trillion-parameter MoE models. Representative examples include the 405B-parameter dense Llama~3.1 model~\cite{Llama3}, the 1T-parameter MoE Kimi~K2 model~\cite{KimiK2}, the 1.6T-parameter DeepSeek-V4-Pro model ~\cite{deepseekv4}, and the 2.8T-parameter Kimi~K3 model~\cite{kimik3}.  For proprietary frontier model families, including OpenAI's GPT, Anthropic's Claude, and Google's Gemini, parameter counts and detailed architectures are generally not publicly disclosed. Third-party analyses nevertheless suggest that selected models may contain hundreds of billions to low trillions of total parameters. For example, estimates for GPT-4 range from approximately 0.4-0.5 trillion parameters under compute-scaling assumptions to about 1.8 trillion total parameters under a mixture-of-experts interpretation~\cite{CSETScalingAI,EpochAIModels}. These estimates depend strongly on assumptions about model architecture, training compute, and the distinction between total and active parameters, and should therefore be interpreted only as rough orders of magnitude rather than directly comparable model sizes.

\begin{table*}[t]
\centering
\small
\caption{Illustrative accelerator-count lower bounds for inference
placement and training Llama~3.1 405B. The inference-memory column
accounts only for BF16 weights. The training-memory column accounts
for fully sharded mixed-precision Adam parameter states. Runtime
memory is excluded from both columns.}
\label{tab:accelerator_instantiation}
\begin{tabular}{lccrrr}
\toprule
Accelerator &
Reported HBM &
BF16 peak &
$N_{\mathrm{mem}}^{\mathrm{infer}}$ &
$N_{\mathrm{mem}}^{\mathrm{train}}$ &
$N_{\mathrm{comp}}^{\mathrm{train}}$ \\
&
&
(PFLOP/s) &
(weights only) &
(training states) &
(60 days) \\
\midrule

NVIDIA H100~\cite{NVIDIA_H100}
    & 80 GB  & 0.989 & 13 & 102 & 18,478 \\
NVIDIA H200~\cite{NVIDIA_H200}
    & 141 GB & 0.989 & 8  & 58  & 18,478 \\
NVIDIA B200~\cite{NVIDIA_HGX_B200}
    & 180 GB & 2.250 & 6  & 45  & 8,125 \\
AMD MI300X~\cite{AMD_MI300X_Platform}
    & 192 GB & 1.307 & 6  & 43  & 13,983 \\
AMD MI355X~\cite{AMD_MI355X}
    & 288 GB & 2.517 & 4  & 29  & 7,265 \\
Ascend 910C~\cite{semianalysis2025huawei,CloudMatrix384}
    & 128 GB & $\sim$0.780 & 8 & 64 & 23,438 \\
Intel Gaudi 3~\cite{Intel_Gaudi3}
    & 128 GB & 1.835 & 8  & 64  & 9,963 \\
Google TPU v5p~\cite{Google_TPU_Machines}
    & 102.0 GB & 0.459 & 10 & 80  & 39,829 \\
Google TPU v6e~\cite{Google_TPU_Machines}
    & 34.4 GB  & 0.918 & 30 & 236 & 19,915 \\
Google TPU7x~\cite{Google_TPU_Machines}
    & 206.2 GB & 2.307 & 5  & 40  & 7,925 \\

\bottomrule
\end{tabular}
\vspace{2pt}
\begin{minipage}{0.98\textwidth}
\footnotesize
\textit{Validation against a reported deployment.}
Meta reports that Llama~3.1 405B was trained using more than 16,000 H100 GPUs~\cite{Llama3}. This reported deployment scale is close to our compute-based estimate of 18,478 H100 GPUs. 
\end{minipage}
\end{table*}

\smalltitle{Distributed execution.}
Distributed LLM execution commonly combines four parallelism dimensions: data parallelism (DP), pipeline parallelism (PP), tensor parallelism (TP), and, for mixture-of-experts models, expert parallelism (EP) as shown in Figure~\ref{fig:parallelism}. DP, PP, and TP are often referred to collectively as three-dimensional parallelism, while EP introduces an additional dimension for distributing experts across accelerators~\cite{rasley2020deepspeed,MetaScale,TPUv4}.

DP replicates a model, or a model shard, across multiple accelerator groups to increase training or serving throughput. During training, DP replicas synchronize gradients or parameter shards through collectives such as AllReduce, ReduceScatter, and AllGather. During inference, replicas can process independent requests without synchronizing model updates~\cite{rasley2020deepspeed,nccl228docs}. PP assigns different ranges of model layers to different accelerator groups. Adjacent pipeline stages exchange activations during the forward pass and activation gradients during the backward pass. Its communication pattern is therefore dominated by point-to-point transfers between neighboring stages~\cite{huang2019gpipeefficienttraininggiant,narayanan2019pipedream}. TP partitions the computation and parameters of individual operators across multiple accelerators. Participating accelerators exchange activations or partial tensor results through frequent collectives such as AllReduce, AllGather, and ReduceScatter. Because these operations lie on the execution critical path, TP is particularly sensitive to communication latency and per-accelerator bandwidth~\cite{MetaScale,nccl228docs}. EP distributes the experts of an MoE layer across different accelerators. Tokens are routed to the accelerators hosting their selected experts and returned after expert computation, producing All-to-All-like dispatch and combine operations. These transfers are often uneven and latency-sensitive because token assignments vary dynamically across experts~\cite{deepseekv32024,lepikhin2020gshard}.



These dimensions solve different capacity and performance problems and communicate different data objects. TP, PP, and EP generally keep their assigned weights resident in accelerator memory and communicate activations or partial results during execution. DP training is the primary case in which communication volume is directly related to replicated parameter state. The selected parallel mapping therefore determines both the required accelerator scale and the communication operations that the interconnect should support.

\subsection{Required Scale-Up Domain Size}
\label{sec:domain_mapping}


\smalltitle{Memory- and compute-capacity bounds.}
The global number of accelerators required by a workload depends jointly on model architecture, numerical precision, runtime state, accelerator capability, and performance objective. Let $P_{\mathrm{total}}$ denote the number of resident model parameters, $b_{\mathrm{state}}$ the average number of bytes required per parameter state, and $M_{\mathrm{runtime}}$ the memory used by activations, KV cache, temporary workspace, and communication buffers. For an accelerator with HBM capacity $M_{\mathrm{HBM}}$, a first-order memory-capacity lower bound is
\[
N_{\mathrm{mem}}
=
\left\lceil
\frac{P_{\mathrm{total}}b_{\mathrm{state}}
      +M_{\mathrm{runtime}}}
     {\eta M_{\mathrm{HBM}}}
\right\rceil ,
\]
where $\eta<1$ is the usable fraction of HBM after runtime reservations, allocator fragmentation, and capacity headroom. Existing LLM runtimes commonly expose configurable usable-memory fractions in the range of approximately $0.8$-$0.9$~\cite{vLLMMemoryDocs,TensorRTLLMKVCache}.

Memory capacity alone is insufficient to determine system scale. Let $C_{\mathrm{work}}$ denote the required computation, $F_{\mathrm{acc}}$ the peak throughput of one accelerator at the relevant precision, $\rho_{\mathrm{comp}}$ the fraction of peak throughput sustained by the workload, and $T_{\mathrm{target}}$ the target execution time. A first-order compute-capacity lower bound is
\[
N_{\mathrm{comp}}
=
\left\lceil
\frac{C_{\mathrm{work}}}
     {\rho_{\mathrm{comp}}F_{\mathrm{acc}}T_{\mathrm{target}}}
\right\rceil .
\]
The resulting global system-scale lower bound is
\[
N_{\mathrm{system}}^{\mathrm{lb}}
=
\max\left(N_{\mathrm{mem}},N_{\mathrm{comp}}\right).
\]
These expressions assume balanced work distribution and ideal scaling. Replicated states, pipeline bubbles, kernel inefficiency, communication overhead, expert imbalance, and synchronization can increase the actual requirement~\cite{ZeRO,jiang2024megascale,TPUv4}.

\smalltitle{Instantiation with existing accelerators.}
The memory- and compute-capacity bounds can be instantiated using publicly reported accelerator specifications. For a dense model trained on $D$ tokens, we approximate the training computation as $C_{\mathrm{train}}\approx6PD$~\cite{kaplan2020scaling,Llama3}. For an MoE model, parameter storage is determined primarily by $P_{\mathrm{total}}$, whereas first-order computation is more closely related to the activated parameter count $P_{\mathrm{act}}$
~\cite{lepikhin2020gshard,deepseekv32024}.

Table~\ref{tab:accelerator_instantiation} illustrates three distinct accelerator-count lower bounds for Llama~3.1 405B. The inference-memory bound assumes BF16 weights at two bytes per parameter and excludes the KV cache and other runtime memory. The training-memory bound assumes an illustrative 16-byte-per-parameter mixed-precision Adam state, including weights, gradients, master parameters, and optimizer states, but excluding activations and temporary workspace. The training-compute bound assumes 15.6 trillion training tokens, $T_{\mathrm{target}}=60$ days, and $\rho_{\mathrm{comp}}=0.4$. All memory estimates assume $\eta=0.8$ usable HBM capacity.

The results demonstrate that memory capacity and computation constrain global system scale differently. The model and optimizer states can in principle be distributed over tens to hundreds of current accelerators, whereas completing the assumed training run within 60 days requires thousands to tens of thousands of accelerators. For example, the estimate requires approximately 18.5K H100 GPUs, which is of the same order as Meta's reported deployment of more than 16K H100 GPUs for Llama~3.1 405B ~\cite{Llama3}. This comparison is a scale-level sanity check rather than a reconstruction of the actual training run.

The interpretation differs for sparse MoE models. Their aggregate HBM requirement is governed by $P_{\mathrm{total}}$, because all experts must be stored, whereas their first-order computation is governed more directly by $P_{\mathrm{act}}$. A trillion-parameter MoE model can therefore require a large model-parallel placement for capacity even when its per-token computation is comparable to that of a much smaller dense model. Expert imbalance, replicated non-expert parameters, and routing buffers can further increase the capacity requirement~\cite{deepseekv32024,deepseekv4}.

\smalltitle{Mapping global scale onto scale-up domains.}
The global accelerator requirement does not imply that all accelerators must reside in one scale-up domain. Instead, we consider a system composed of multiple model-complete domain-level execution meshes. Each mesh contains the PP, TP, EP, and any required in-domain state-sharding placements needed to execute one copy of the model. Let $N_{\mathrm{map}}$ denote the accelerator count of one such model-complete mesh, and let $N_{\mathrm{mesh}}$ denote the number of meshes deployed globally. The resulting global accelerator count is
\[
N_{\mathrm{global}}
=
N_{\mathrm{mesh}}N_{\mathrm{map}}
\ge
N_{\mathrm{system}}^{\mathrm{lb}},
\]
and the minimum number of meshes is
\[
N_{\mathrm{mesh}}
=
\left\lceil
\frac{N_{\mathrm{system}}^{\mathrm{lb}}}
     {N_{\mathrm{map}}}
\right\rceil.
\]

\begin{figure}
    \centering
    \includegraphics[width=1\linewidth]{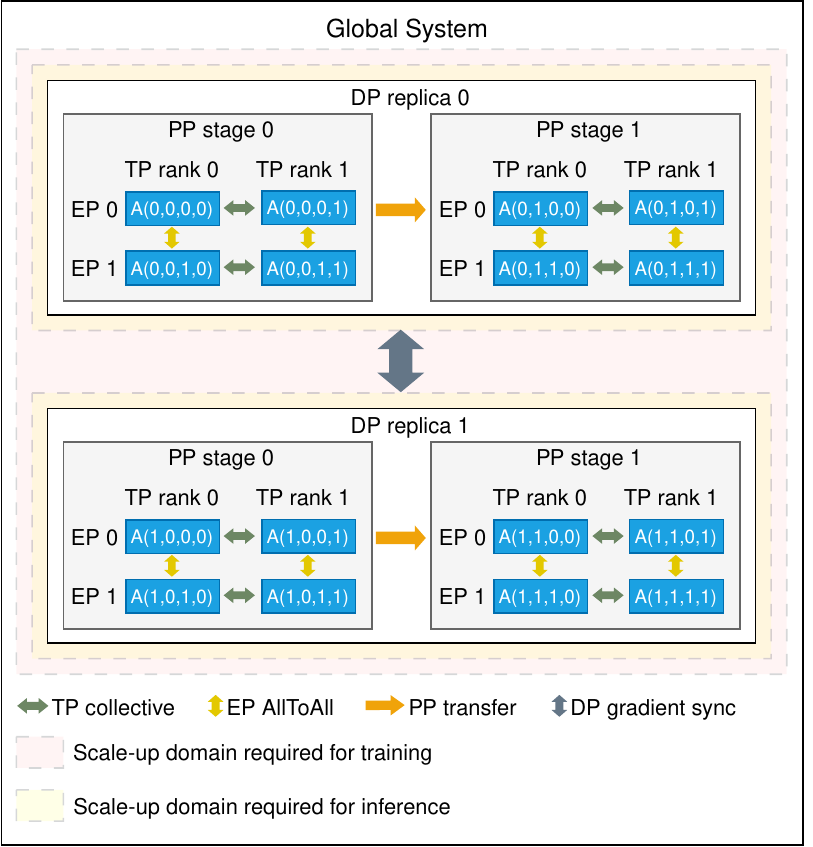}
    \caption{Parallelism Mapping onto Scale-Up Domains. The illustrated global system contains an orthogonal \(DP2\times PP2\times EP2\times TP2\) mapping over 16 accelerators. During inference, each model-parallel replica forms an independent eight-accelerator domain. During training, the two replicas may be colocated within a 16-accelerator domain when DP gradient synchronization remains on the scale-up fabric. The colored boundaries represent workload-derived placement requirements rather than changes to the physical interconnect. Each accelerator is indexed as \(A(d,p,e,t)\), where the four coordinates denote its DP, PP, EP, and TP ranks, respectively.}
    \label{fig:parallelism-domain-mapping}
\end{figure}

The interpretation of $N_{\mathrm{mesh}}$ depends on the workload. During synchronous training, the domain-level meshes may participate in cross-domain data parallelism and synchronize gradients through hierarchical collectives. In this case, $N_{\mathrm{mesh}}$ corresponds to the cross-domain DP degree. During inference, however, the meshes are generally independent serving replicas whose number is determined by request throughput, latency objectives, locality, and fault isolation. They do not normally communicate on the per-token critical path. Figure ~\ref{fig:parallelism-domain-mapping} illustrates how the same global parallelism mapping can induce different scale-up-domain requirements for training and inference. We therefore use $N_{\mathrm{mesh}}$ as the general deployment count rather than treating it as a DP degree in all workloads.

During synchronous training, a hierarchical collective can first aggregate gradients or parameter shards within each scale-up domain, perform inter-domain synchronization across the $N_{\mathrm{mesh}}$ domain-level meshes through the scale-out network, and then distribute the result within each domain. During inference, the same decomposition allows independent serving replicas to be placed in different domains without introducing per-token cross-domain synchronization.

\smalltitle{Illustrative mapping instantiation.}
We next instantiate the mapping model using representative configurations reported for dense training, dense inference, and MoE execution. Table~\ref{tab:domain_size_instantiation} summarizes the resulting model-complete placement sizes and their mapping basis.

\begin{table}[t]
\centering
\scriptsize
\caption{Representative model mappings relevant to scale-up-domain sizing. ''Derived'' denotes a placement size calculated from publicly reported parallelism degrees; it does not imply that all accelerators were physically connected by one scale-up fabric.}
\label{tab:domain_size_instantiation}
\begin{tabular}{
@{}p{1.45cm}
p{2.50cm}
p{1.55cm}
p{1.75cm}@{}
}
\toprule
Workload &
Mapping configuration &
Placement size &
Basis \\
\midrule

Llama~3.1 training &
TP8 $\times$ PP16
&
$N_{\mathrm{map}}=128$
&
Derived from reported mapping~\cite{Llama3} \\

\addlinespace

Llama~3.1 inference &
TP8 within each node and PP2 across two nodes
&
$N_{\mathrm{map}}=16$
&
Reported deployment~\cite{Llama3} \\

\addlinespace

DeepSeek-V3 training &
TP1, PP16, and EP64 with ZeRO-1 DP
&
$N_{\mathrm{map}}=1{,}024^{*}$
&
Derived from reported mapping~\cite{deepseekv32024} \\

\addlinespace

DeepSeek-V3 prefill &
Attention: TP4 $\times$ DP8;
MoE: EP32
&
$N_{\mathrm{map}}=32$
&
Reported minimum unit~\cite{deepseekv32024} \\

\bottomrule
\end{tabular}

\vspace{2pt}

\begin{minipage}{0.96\columnwidth}
\tiny
$^{*}$The DeepSeek-V3 training report states that EP64 spans eight nodes and that the full training cluster contains 2,048 H800 GPUs. The 1,024-accelerator value is therefore a model-parallel footprint, not evidence that all 1,024 accelerators form one physical scale-up domain.
\end{minipage}
\end{table}

\begin{table*}[t]
\centering
\small
\caption{Communication generated by parallel LLM execution. Payloads are leading-order logical volumes per participating accelerator.}
\label{tab:parallel_comm_req}
\begin{tabular}{p{0.07\textwidth} p{0.18\textwidth}
                p{0.21\textwidth} p{0.18\textwidth}
                p{0.27\textwidth}}
\toprule
\textbf{Dimension} &
\textbf{Typical operation} &
\textbf{Communicated data} &
\textbf{Logical payload} &
\textbf{Performance characteristics} \\
\midrule

DP &
AllReduce or ReduceScatter--AllGather &
Gradients or parameter shards &
$\Theta(P_gb_g)$ per synchronization &
In-domain sharded DP requires high sustained bandwidth. Any non-overlapped communication tail delays the completion of the training iteration~\cite{ZeRO,rasley2020deepspeed,nccl228docs}. \\

PP &
Point-to-point transfer &
Forward activations or backward activation gradients &
$\Theta(THb_a)$ per stage boundary &
The receiving stage cannot proceed until the transfer completes; exposed communication latency increases pipeline bubbles
~\cite{huang2019gpipeefficienttraininggiant,narayanan2019pipedream}. \\

TP &
AllReduce, AllGather, or ReduceScatter &
Activations or partial operator results &
$\Theta(THb_a)$ per collective &
Collectives are invoked repeatedly within Transformer layers and therefore require both high effective bandwidth and low completion latency~\cite{shoeybi2019megatronlm,nccl228docs}. \\

EP &
All-to-All-like dispatch and combine &
Token representations sent to and returned from experts &
$\Theta(f_{\mathrm{remote}}TkHb_a)$ per phase &
Dispatch and combine surround expert computation. Routing imbalance can produce bursty and nonuniform traffic
~\cite{lepikhin2020gshard,deepseekv32024}. \\

\bottomrule
\end{tabular}
\end{table*}

We select a candidate $N_{\mathrm{map}}$ using three criteria. First, the placement must provide sufficient aggregate memory for the relevant model state. Second, its TP, PP and EP degrees must be compatible with the model structure and use regular group sizes. Third, communication-intensive dimensions, particularly TP and EP, should remain within the lowest-latency communication domain whenever possible.

For Llama~3.1 405B training, Meta reports TP8 and PP16 for its 8K-token pretraining configurations~\cite{Llama3}. Their product gives a 128-accelerator model-parallel placement, which also exceeds the $N_{\mathrm{mem}}^{\mathrm{train}}=102$ memory bound derived in Table~\ref{tab:accelerator_instantiation}. We therefore use $N_{\mathrm{map}}=128$ rather than constructing an arbitrary factorization of the memory bound. With $N_{\mathrm{comp}}^{\mathrm{train}}=18{,}478$, the illustrative global requirement becomes $ N_{\mathrm{mesh}} = \left\lceil \frac{18{,}478}{128} \right\rceil = 145, \qquad N_{\mathrm{global}} = 145\times128 = 18{,}560$.

For BF16 inference, Meta reports that Llama~3.1 405B is deployed on 16 H100 GPUs using TP8 within each eight-GPU node and PP2 across two nodes~\cite{Llama3}. This configuration satisfies the $N_{\mathrm{mem}}^{\mathrm{infer}}=13$ weight-placement lower bound and provides a publicly reported value of $N_{\mathrm{map}}=16$ per serving replica.

The DeepSeek-V3 configurations illustrate that MoE mappings need not be simple products of the same parallel dimensions used for dense models. Its training configuration uses PP16 and EP64 without TP, while its prefilling deployment uses different mappings for the attention and MoE components~\cite{deepseekv32024}. These examples also show why a reported global mapping should not automatically be interpreted as one physical scale-up domain: some communication groups explicitly span multiple nodes over the cluster network.


\subsection{Communication Requirements within a Scale-Up Domain}
\label{sec:communication_requirements}

The placement scale derived in the preceding section determines the accelerator coverage required when the corresponding mapping is intended to remain within one scale-up domain. It does not by itself determine the required network performance.

Let $T$ denote the number of token representations processed by a communication group in one microbatch, $H$ the hidden dimension, $b_a$ the number of bytes per activation element, and $P_g$ the number of parameter elements synchronized by a DP group. Table~\ref{tab:parallel_comm_req} summarizes the leading-order logical payload generated by each parallel dimension. The payload is expressed per participating accelerator before accounting for the communication algorithm and protocol overhead.

Here, $b_g$ is the number of bytes per synchronized gradient or parameter element, $k$ is the number of experts selected for each token, and $f_{\mathrm{remote}}$ is the fraction of expert assignments sent to nonlocal accelerators. The values in Table~\ref{tab:parallel_comm_req} are logical payloads rather than physical wire traffic. Actual traffic additionally depends on the collective algorithm, group size, topology, routing, packet overhead, and retransmission behavior. For example, a ring AllReduce over $n$ accelerators transfers approximately $2(n-1)/n$ times the input tensor size at each accelerator~\cite{nccl228docs,taccl}.

\smalltitle{Bandwidth and latency requirements.}
The completion time of a communication operation $g$ can be approximated using a latency-bandwidth model~\cite{dally2004interconnection,nccl228docs}:
\[
T_{\mathrm{comm},g}
\approx
r_g\alpha_g+
\frac{D_{\mathrm{wire},g}}{B_{\mathrm{eff},g}},
\]
where $r_g$ is the number of serialized communication rounds, $\alpha_g$ is the effective latency of each round, $D_{\mathrm{wire},g}$ is the algorithm-dependent wire traffic, and $B_{\mathrm{eff},g}$ is the effective bandwidth available to the communication group. Here, $\alpha_g$ captures the size-independent delay of each serialized round, including collective launch, endpoint processing,
switching, and synchronization overhead, whereas $B_{\mathrm{eff},g}$ denotes the achieved data-transfer rate after accounting for topology, routing, protocol overhead, and contention.

Training and long-context prefill generally exchange large gradients, activations, or routed token tensors and are therefore sensitive to sustained bandwidth~\cite{jiang2024megascale,rasley2020deepspeed}. Autoregressive decoding exchanges smaller tensors, but invokes TP and EP communication repeatedly across many layers for every generated token. Its critical path is consequently more sensitive to collective launch, synchronization, endpoint processing, switching, and queueing latency~\cite{DistServe,deepseekv32024}.

\smalltitle{Illustrative communication-requirement instantiation.}
The following examples instantiate the latency-bandwidth model for three representative communication patterns: bandwidth-dominated TP, latency-sensitive decoding, and bandwidth- and balance-sensitive EP. Table~\ref{tab:communication_instantiation} summarizes the resulting performance requirements. These values describe effective one-way performance at the accelerator interface and should only be interpreted as representative orders of magnitude.

\begin{table}[t]
\centering
\small
\setlength{\tabcolsep}{3pt}
\caption{Summary of the illustrative communication requirements derived from the representative TP, decoding, and EP workloads.}
\label{tab:communication_instantiation}
\begin{tabular}{@{}p{2.4cm}p{5.0cm}@{}}
\toprule
Scenario &
Derived requirement \\
\midrule

Bandwidth-dominated TP &
235-470 GB/s of effective one-way bandwidth per accelerator \\

\addlinespace

Latency-sensitive decode &
4-10 $\mu$s average completion-time budget per serialized collective \\

\addlinespace

Bandwidth- and balance-sensitive EP &
117-235 GB/s per accelerator and 7.5-15 TB/s of aggregate endpoint injection bandwidth for a 64-accelerator EP group \\

\bottomrule
\end{tabular}
\end{table}

\emph{Bandwidth-dominated TP.}
Consider Llama~3.1 405B, with hidden dimension $H=16{,}384$, and a TP group processing $T=8{,}192$ BF16 token representations~\cite{Llama3}. The corresponding activation tensor is approximately 268.4~MB. For an eight-accelerator ring AllReduce, each accelerator injects approximately 469.8~MB of data. Completing this collective within 2~ms requires approximately 235~GB/s of effective one-way bandwidth per accelerator, while reducing the communication budget to 1~ms increases the requirement to approximately 470~GB/s. 
For the same prefill configuration, a PP stage boundary transfers one 268.4~MB activation tensor. Completing this point-to-point transfer within 1--2~ms requires approximately 134-268~GB/s, which is lower than the 235-470~GB/s required by the eight-accelerator TP collective. During decode, PP transfers one token representation across each stage boundary, whereas TP collectives are repeatedly invoked within every Transformer layer. TP therefore determines both $B_{\mathrm{req}}$ and $L_{\mathrm{req}}$ for this illustrative mapping.

\emph{Latency-sensitive decode.}
Llama~3.1 405B contains 126 layers and has a hidden dimension of 16,384~\cite{Llama3}. A single BF16 token representation therefore contains only 32.8~KB, making serialization bandwidth less important than fixed communication overhead. As an illustrative design point, assume that the TP implementation invokes two serialized collectives per layer, the target time per output token (TPOT) is 10-25~ms, and exposed TP communication may consume at most 10\% of this budget. The average completion-time budget for each collective is then only approximately 4-10~$\mu$s. Although the precise value depends on operator partitioning, batching, kernel fusion, and communication-computation overlap, this example shows that aggressive TPOT targets can leave only single-digit to low-tens-of-microseconds for each serialized collective~\cite{DistServe,nccl228docs}.

\emph{Bandwidth- and balance-sensitive EP.}
DeepSeek-V3 has hidden dimension $H=7{,}168$ and selects eight of 256 routed experts for each token~\cite{deepseekv32024}. Suppose that one accelerator dispatches $T=512$ BF16 token representations and, conservatively, that all selected experts are remote. The dispatch payload is approximately 58.7~MB, and the combine phase returns a similar amount. Assigning the resulting 117.4~MB of communication an exposed budget of 0.5-1~ms requires approximately 117-235~GB/s of effective one-way bandwidth per accelerator. Across a 64-accelerator EP group, this corresponds to approximately 7.5-15~TB/s of aggregate endpoint injection bandwidth. Expert locality can reduce this traffic, whereas routing imbalance can produce substantially higher local hotspots~\cite{lepikhin2020gshard,deepseekv32024}.

Together, these examples suggest an illustrative performance envelope of approximately 100-500~GB/s of effective per-accelerator bandwidth for bandwidth-intensive TP and EP communication, and a few to tens of microseconds of collective completion time for latency-sensitive decoding. The precise requirements must nevertheless be derived from the model architecture, numerical precision, parallel placement, collective algorithm, and application-level performance objective.

\smalltitle{Concurrency and performance isolation.}
The preceding requirements describe a workload in isolation. In a shared domain, concurrent DP, PP, TP, EP, and multi-tenant traffic may contend for links, buffers, routing paths, and collective resources, reducing effective bandwidth and increasing completion time. A practical scale-up fabric must therefore preserve the derived requirements under its admitted load. This observation motivates the resource-partitioning, admission-control, and performance-isolation challenges discussed in Section~\ref{subsec:resource-sharing}.


\subsection{Workload-Derived Requirements}
\label{sec:scaleup_definition}

Following public specifications and system documentation, we use the term \emph{scale-up domain} to denote a fabric-scoped set of accelerator endpoints that are mutually reachable under a common set of addressing, routing, communication, and management mechanisms
~\cite{NVLink,UALink_spec,SUE_Whitepaper}. Public descriptions typically characterize such domains through fabric membership, endpoint scale, connectivity, and qualitative performance objectives such as high bandwidth and low latency.

Such a definition identifies which accelerators belong to the same domain, but does not determine the domain capabilities required by a particular workload. We therefore summarize the preceding placement and communication analyses as three workload-derived requirements: scale, bandwidth, and latency.

For a workload $W$ under parallel mapping $\mathcal{M}$, we represent these requirements as
\[
\mathcal{R}(W,\mathcal{M})
=
\left(
S_{\mathrm{req}}(\mathcal{M}),
B_{\mathrm{req}},
L_{\mathrm{req}}
\right).
\]
Here, $S_{\mathrm{req}}(\mathcal{M}) = |\mathcal{V}_{\mathcal{M}}| = N_{\mathrm{map}}$ is the minimum number of mutually reachable accelerators required to contain the model-complete execution mesh derived in Section~2.2. $B_{\mathrm{req}}$ is the effective per-accelerator bandwidth required by the most bandwidth-intensive communication phase, while $L_{\mathrm{req}}$ is an end-to-end collective completion-time requirement, rather than a per-link latency target, derived in Section~2.3. The bandwidth and latency requirements may originate from different execution phases and should therefore be interpreted as complementary constraints rather than as two estimates of the same communication operation.

Table~\ref{tab:scaleup_requirement_summary} instantiates this requirement tuple for the Llama~3.1 405B inference example used above.
Using the reported $PP2\times TP8$ arrangement as the selected model-complete in-domain mapping, we obtain $S_{\mathrm{req}}(\mathcal{M})=16$ accelerators. This is a workload-driven placement choice; the reported deployment does not by itself establish that all 16 accelerators belonged to one physical scale-up domain.
The bandwidth-dominated tensor-parallel phase requires approximately 235-470~GB/s of effective one-way bandwidth per accelerator, whereas latency-sensitive decode requires serialized collectives to complete within approximately 4-10~$\mu$s. These values are taken from the placement and communication instantiations in Sections~2.2 and~2.3, respectively.

\begin{table}[t]
\centering
\caption{Workload-derived scale-up requirements for the illustrative Llama~3.1 405B inference configuration.}
\label{tab:scaleup_requirement_summary}
\begin{tabular}{l c}
\toprule
\textbf{Requirement} & \textbf{Derived value} \\
\midrule
Scale, $S_{\mathrm{req}}$
    & 16 accelerators \\
Effective bandwidth, $B_{\mathrm{req}}$
    & 235-470 GB/s per accelerator \\
Collective completion time, $L_{\mathrm{req}}$
    & 4-10 $\mu$s \\
\bottomrule
\end{tabular}
\end{table}

Accordingly, a candidate scale-up domain can support this workload only if it contains at least $S_{\mathrm{req}}(\mathcal{M})$ mutually reachable accelerators, provides at least $B_{\mathrm{req}}$ effective bandwidth during bandwidth-intensive phases, and satisfies $L_{\mathrm{req}}$ for latency-sensitive collectives under the specified load. The scale value is a workload-derived lower bound, rather than a prescribed physical domain size: a larger domain remains suitable provided that its topology and communication performance satisfy the same requirements.

The workload-derived domain size should, however, be distinguished from the maximum scale supported by a particular fabric implementation. The former is determined by model placement and communication requirements, whereas the latter is additionally constrained by the fabric’s addressing and identifier space, routing state, topology, and management architecture. These implementation-side constraints are examined in Section~\ref{sec:3} through our analysis of topology and scale, communication and transport, and system integration.

\smalltitle{Distinguishing interconnect regimes.}
Under this workload-dependent definition, scale-up fabrics should be distinguished from NoC/scale-in fabrics and scale-out networks by the communication regime they provide, rather than by a fixed endpoint count, physical-link rate, or implementation technology.

NoC and scale-in fabrics are designed to provide high-bandwidth, low-latency communication within a tightly integrated local system. Representative implementations provide hundreds of Gb/s per link to tens of TB/s of aggregate or cross-sectional bandwidth, with communication latency ranging from a few to tens of nanoseconds. Their scale ranges from a small number of chiplets or accelerators within a package or board to hundreds or thousands of on-chip processing elements~\cite{FlooNoC,KurthNoC,UCIe2Spec}. NoC/scale-in fabrics therefore occupy the high-bandwidth and low-latency region of the design space, but remain confined to a chip, package, board, or other local system boundary.


Scale-out networks make a different trade-off. They are designed to connect thousands or more endpoints across servers and racks. Current high-performance network adapters reach 800~Gb/s, or 100~GB/s per port~\cite{connectx8}, but this bandwidth may be shared among multiple accelerators. Communication also typically traverses multiple switching stages and potentially oversubscribed links, while network-wide congestion and queuing further reduce sustained bandwidth and increase completion-time variability~\cite{swift,HPCC,dcqcn}. Scale-out networks can therefore provide the required global deployment scale, but cannot efficiently provide the high and predictable per-accelerator bandwidth required by communication-critical TP and EP groups.

Scale-up fabrics occupy a workload-defined regime between these two design objectives. Like NoC/scale-in fabrics, they are intended to support tightly coupled accelerator communication with high effective bandwidth and low, predictable completion latency. Unlike local scale-in fabrics, however, they must extend these properties to a sufficiently large accelerator group to contain the communication-intensive dimensions of a distributed workload. For workload $W$ under parallel mapping $\mathcal{M}$, a candidate fabric qualifies as scale-up only if it connects at least $S_{\mathrm{req}}(\mathcal{M})$ mutually reachable accelerators, sustains the required effective per-accelerator bandwidth $B_{\mathrm{req}}$, and satisfies the collective-completion-time requirement $L_{\mathrm{req}}$ under the specified load. In the illustrative workloads analyzed above, this regime corresponds to accelerator groups ranging from tens to potentially hundreds or more participants, effective per-accelerator bandwidth on the order of hundreds of GB/s, and collective completion times ranging from a few to tens of microseconds.

\begin{table*}[t]
\centering
\scriptsize
\caption{Background and public evidence for the eleven scale-up fabric families, with SUE and SUE-T shown separately as related specifications.}

\label{tab:info}
\setlength{\tabcolsep}{2.6pt}
\renewcommand{\arraystretch}{1.15}

\begin{tabularx}{\textwidth}{
@{}
>{\raggedright\arraybackslash}p{0.095\textwidth}
>{\raggedright\arraybackslash}p{0.095\textwidth}
>{\raggedright\arraybackslash}p{0.135\textwidth}
>{\raggedright\arraybackslash}p{0.130\textwidth}
>{\raggedright\arraybackslash}X
>{\raggedright\arraybackslash}p{0.115\textwidth}
>{\raggedright\arraybackslash}p{0.110\textwidth}
@{}}

\toprule

\textbf{Fabric Family} &
\textbf{Developer / Governance} &
\textbf{Public Specification Status$^*$} &
\textbf{Technical Lineage} &
\textbf{Fabric Composition and Scope} &
\textbf{Reference Public Milestone$^\dagger$} &
\textbf{Evidence Maturity} \\

\midrule

NVLink/ NVSwitch
\cite{NVLink,Nvidia_NVSwitch} &
NVIDIA &
Vendor-controlled; public product and programming documentation &
Purpose-built accelerator interconnect &
GPU endpoint interfaces, direct NVLink connections, NVSwitch switching ASICs, and centralized fabric-management software &
2014 announcement; 2016 first product generation &
Production-deployed \\

\addlinespace[2pt]

Infinity Fabric
\cite{InfinityFabric} &
AMD &
Vendor-controlled; public product and architectural documentation &
Coherent processor and accelerator interconnect &
On-package and board-level links, I/O dies, and direct GPU-to-GPU Infinity Fabric links & 2020 AMD Instinct MI100 generation\footnotemark[1] &
Production-deployed \\

\addlinespace[2pt]

ICI
\cite{Google_TPU8,Jouppi2023TPUv4} &
Google &
Vendor-controlled; architecture described in public system papers &
Purpose-built TPU interconnect &
TPU endpoint links, packet-switched interconnects, torus topologies, and optical circuit switches in selected generations &
2017 TPU v2 public disclosure &
Production-deployed \\

\addlinespace[2pt]

ICN
\cite{THeadICN,THeadICNSwitch} &
T-Head &
Vendor-controlled; limited public protocol documentation &
Purpose-built accelerator interconnect &
Zhenwu accelerator interfaces, proprietary ICN links, ICN Switch 1.0, and the PCCL communication stack &
May 2026 product launch &
Product released; protocol details limited \\

\addlinespace[2pt]

Unified Bus
\cite{UBSpec,openeuler2025unifiedbus} &
UnifiedBus community; initially led by Huawei &
Publicly downloadable specifications under applicable licensing terms &
Unified device and system interconnect; not classified solely as Ethernet-derived$^{**}$ &
Endpoint and I/O interfaces, direct links, UB switches, firmware, management software, and programming models &
Sept.\ 2025 public launch; Sept.\ 2026 Base Spec.\ 2.1 &
Public specification; vendor-reported production deployment \\

\addlinespace[2pt]

SUE
\cite{SUE_Whitepaper,SUE} &
Broadcom; contributed to OCP &
Public vendor-authored OCP specification &
Ethernet-derived endpoint transaction and transport protocol &
XPU-facing transaction and transport functions, link-level reliability and flow control mechanisms, and interfaces to an Ethernet link and switch fabric; provides a lightweight SUE-Lite variant without transport-layer functions. &
Apr.\ 2025 framework; Sept.\ 2025 OCP 1.0 &
Public specification; supporting products available \\

\addlinespace[2pt]

SUE-T~\cite{SUE_T_Base_1_0} &
OCP workstream &
Public SUE-T Base Specification 1.0.0 &
Derived from SUE submitted by Broadcom to OCP; uses UEC Unified Local/RODL over ESUN, UFH, or UDP/IP/Ethernet &
XPU transaction mapping, packing, scheduling, and endpoint transport; lower-layer network services supplied by the selected Ethernet fabric &
Oct. 2025 rename; Oct. 2026 base specification &
Public specification \\

\addlinespace[2pt]

UALink
\cite{UALink_Common_2_0,UALink_200G_DLPL_2_0,UALink_128G_DLPL_1_0, UALink_spec,UALink,UALink_Webinar} &
UALink Consortium &
Public consortium specifications distributed under consortium terms &
Common transaction architecture with an IEEE~802.3-derived 200G profile and a PCIe-derived 128G profile &
Accelerator interfaces, direct links, retimers, UALink switches, and centralized management components. UALink Common 2.0 additionally introduces in-network-compute capabilities, while the data-link and physical-layer specifications are maintained as separately versioned documents. &
May 2024 initiative; Apr. 2025 UALink 200G 1.0; Apr. 2026 Common 2.0 and 200G DL/PL 2.0 &
Public specifications available; product and interoperability ecosystem emerging \\

\addlinespace[2pt]

TTP/TTPoE
\cite{teslamotors2024ttpoe,Quinnell2024ttpoe} &
Tesla &
Public technical disclosure; no complete implementer specification identified &
Purpose-built TTP transport mapped onto standard Ethernet MAC/PHY components in TTPoE &
Hardware TTP transport engines, Ethernet-II frames, standard Ethernet MAC/PCS/PHY components, NICs, and Ethernet switches &
Aug.\ 2024 Hot Chips disclosure &
Production system publicly described \\

\addlinespace[2pt]

EthLink~\cite{EthLinkSpec} &
ByteDance &
Public architecture white paper; implementation vendor-controlled &
Ethernet-derived scale-up interconnect &
XPU interfaces, transaction and reliability layers, optimized EthLink forwarding headers, intra-server links, and scale-up switches &
May 2025 white-paper disclosure &
Architecture published; vendor-reported deployment validation \\

\addlinespace[2pt]

ESUN
\cite{OCP_ESUN_Launch,OCPESUN2026} &
Open Compute Project &
Public open-community specification &
Ethernet-derived network and data-link architecture &
Endpoint network controllers, compact ESUN headers, Ethernet links, and single- or multi-tier Ethernet switches &
Oct.\ 2025 workstream; Mar.\ 2026 Spec.\ 1.0 &
Public specification; implementation ecosystem emerging \\

\addlinespace[2pt]

ETH+
\cite{ETHPlusSpec,ETHPlusConsortium} &
ETH+ Consortium, co-led by ICT CAS and Alibaba Cloud &
Public consortium specification and selected open hardware/IP artifacts &
Ethernet-derived identifier-based interconnect &
Endpoint and semantic-adaptation interfaces, optimized Ethernet framing, switches, NIC/IP implementations, and copper or optical links &
July 2023 consortium; Sept.\ 2024 Spec.\ 1.0 &
Public specification; prototypes and system deployments reported \\

\bottomrule
\end{tabularx}

\vspace{2pt}

\begin{minipage}{0.96\linewidth}

\raggedright
\tiny
$^*$ Public specification status describes only the availability of technical documentation and does not imply existing hardware support, unrestricted licensing, or multi-vendor interoperability. 

$^{**}$ ``Ethernet-derived'' indicates only reuse or modification of selected Ethernet or IEEE~802.3 mechanisms rather than direct compatibility with standard Ethernet.

$^\dagger$ Milestone dates refer to the specific events stated in the table and are not treated as directly equivalent product-release dates.
\end{minipage}

\end{table*}

\footnotetext[1]{Infinity Fabric predates the AMD Instinct MI100. The listed date identifies its public use as a direct multi-accelerator interconnect rather than the first use of Infinity Fabric anywhere in an AMD system.}

\begin{table*}[h!tbp]
\centering
\small
\caption{Structural properties and scaling constraints of representative
scale-up topology classes. Latency and bandwidth are not ranked because
they depend on link rate, switch implementation, routing, oversubscription,
and workload traffic patterns.}
\label{tab:topology_properties}
\setlength{\tabcolsep}{3.5pt}
\begin{tabularx}{\textwidth}{
    >{\raggedright\arraybackslash}p{2.2cm}
    >{\raggedright\arraybackslash}p{3.0cm}
    >{\raggedright\arraybackslash}p{2.2cm}
    >{\raggedright\arraybackslash}p{2.3cm}
    >{\raggedright\arraybackslash}X}
\toprule
\textbf{Topology}
& \textbf{Typical Endpoint-to-Endpoint Path}
& \textbf{Endpoint Degree}
& \textbf{Scaling Cost}
& \textbf{Primary Scaling Constraint} \\
\midrule

Full mesh
& One direct link; no external switch traversal
& $N-1$
& $O(N^2)$ links
& Endpoint SerDes count, package pins, board routing, cable density, and endpoint injection bandwidth \\

Single-tier switched
& Endpoint-switch-endpoint within one fabric plane;
  one logical switching tier
& Usually constant, or proportional to the number of
  independent planes
& $O(N)$ endpoint links per plane
& Endpoint-facing switch radix, switch fabric capacity, number of parallel planes, and physical reach \\

Two-tier switched
& Endpoint-leaf-spine-leaf-endpoint for endpoints
  attached to different leaf switches
& Usually constant, or proportional to the number of rails
& $O(N)$ endpoint links plus leaf--spine links
& Leaf and spine radix, uplink oversubscription, inter-tier bandwidth, routing, and path diversity \\

$k$-D mesh/torus
& Multiple nearest-neighbor hops; path length grows with
  topology dimensions
& At most $2k$
& $O(N)$ links
& Network diameter, bisection bandwidth, routing, traffic locality, and hotspot formation \\

Hybrid or hierarchical
& Depends on the combination of direct links, switches, and topology levels
& Implementation-dependent
& Implementation-dependent
& Depends on the placement of hierarchy boundaries, inter-level bandwidth, and workload mapping \\

\bottomrule
\end{tabularx}

\vspace{1cm}

\centering
\caption{Topology support and evidence status of the surveyed scale-up fabrics.}
\label{tab:topo2}
\small
\setlength{\tabcolsep}{4pt}
\begin{tabularx}{\textwidth}{
    >{\raggedright\arraybackslash}p{2.5cm}
    >{\centering\arraybackslash}p{1.2cm}
    >{\centering\arraybackslash}p{1.5cm}
    >{\centering\arraybackslash}p{1.5cm}
    >{\centering\arraybackslash}p{1.2cm}
    >{\raggedright\arraybackslash}X}
\toprule
\textbf{Fabric Family}
& \textbf{Mesh}
& \textbf{Single Tier}
& \textbf{Two Tier}
& \textbf{Others}
& \textbf{Representative Configuration or Qualification} \\
\midrule

NVLink / NVSwitch
& D
& D
& D
& NS
& Hybrid cube mesh in DGX-1; single-stage switched all-to-all in NVL72; two-layer switched all-to-all in the announced Vera Rubin Ultra NVL576 \\

Infinity Fabric
& D
& NS
& NS
& NS
& Fully connected eight-GPU mesh in the MI300X/MI325X platform \\

ICI
& D
& NS
& NS
& D
& Three-dimensional mesh in TPU~v4 and three-dimensional torus in TPU7x/Ironwood; Ironwood scales to 9,216 chips \\

Unified Bus 2.0
& D
& D
& D
& D
& Public documentation describes direct, mesh, and switched organizations. UB-Mesh provides a hierarchical $n$D-full-mesh design, while the supported scale and physical hierarchy remain implementation- and generation-specific. \\

SUE
& D
& D
& NS
& NS
& Public documentation explicitly illustrates direct mesh and single-switch-hop deployments and permits up to 1,024 accelerators per scale-up domain. \\

SUE-T\cite{SUE_T_Base_1_0}
& P
& D
& D
& P
& Endpoint transport specification optimized for single-hop fabrics, with multi-hop extensibility and an address/transport scope of up to 16K accelerators; this figure is not a validated topology scale. Multi-tier congestion control and integrated failover remain future work. \\

UALink
& D
& D
& U
& NS
& Public documentation supports direct connections and single-tier switched pods of up to 1,024 accelerators, while explicitly excluding switch-to-switch links and therefore multi-tier topologies. \\

TTP / TTPoE
& NS
& P
& P
& P
& Supports direct or Ethernet-switched communication, but no single mandatory topology is publicly specified \\

ETH+
& NS
& P
& D
& P
& Documented two-tier UPN~512 architecture; hierarchy is implementation-specific~\cite{ETHPlusUPN} \\

EthLink
& NS
& P
& P
& P
& Ethernet-derived switched fabric; public information does not mandate a unique switch hierarchy \\

ICN
& NS
& P
& P
& P
& Vendor-integrated switched fabric with implementation-dependent topology \\

ESUN
& NS
& P
& P
& P
& Defines network- and switch-side requirements rather than one complete mandatory topology \\

\bottomrule
\end{tabularx}

\vspace{2pt}

\begin{minipage}{0.96\linewidth}
\small
\begin{center}
    \begin{tabular}{@{}l@{}}
    D: explicitly documented in public specifications or implemented in real-world systems; \\
    P: permitted in principle, but lacks explicit statements from public specifications or identified real-world deployment;\\
    U: explicitly unsupported by the cited specification or system profile. \\
    NS: insufficient public information.
  \end{tabular}
\end{center}

\end{minipage}

\vspace{3pt}


\end{table*}

\section{Analysis}
\label{sec:3}
We analyze eleven representative scale-up fabric families with different properties. NVLink/NVSwitch, AMD Infinity Fabric, Google ICI, and T-Head ICN are vertically integrated with particular accelerator platforms and system architectures. Unified Bus 2.0, SUE, SUE-T (an OCP workstream evolved from SUE; We treat SUE and SUE-T as a single scale-up family), UALink, TTP/TTPoE, EthLink, ESUN, and ETH+ are vendor- or consortium-led specifications intended to support broader component or protocol ecosystems, although their levels of openness, interoperability, and deployment maturity differ substantially. Several of these fabrics reuse selected IEEE 802.3 technologies, such as SerDes, PHY, PCS, or Ethernet framing conventions. We refer to such designs as \emph{IEEE 802.3-derived} rather than \emph{Ethernet-compatible}, because reuse of the physical layer does not imply interoperability at the data-link, network, transport, or transaction layers. Table~\ref{tab:info} therefore characterizes each fabric along four independent dimensions: integration model, specification availability, technical lineage, and evidence maturity.


We compare these fabrics along two architectural dimensions and two deployment dimensions. Topology determines path characteristics, available bandwidth, and domain scale; protocol covers data transfer, flow control, reliability, and transport; and semantic execution describes the communication semantics exposed to accelerators and their execution models. Resource sharing and integration considers logical partitioning, access and performance isolation, and heterogeneous endpoint support. Together, these dimensions reveal how existing fabrics trade off performance, scalability, programmability, isolation, and interoperability.

\begin{table*}
    \centering
    \includegraphics[width=0.8\linewidth]{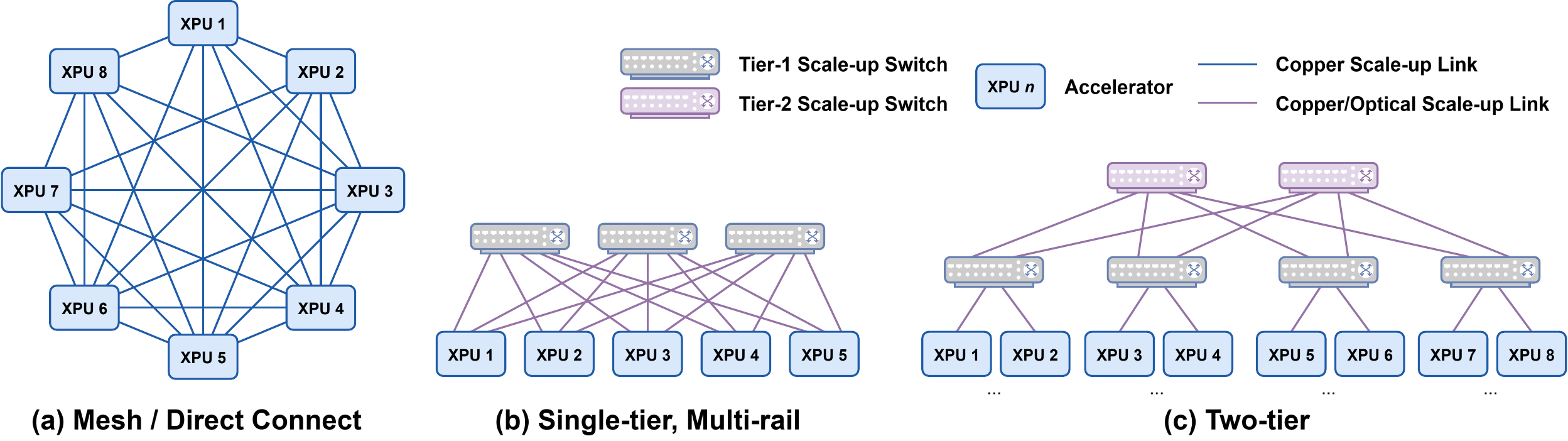}
    \captionof{figure}{Representative scale-up topologies: (a) direct full mesh, (b) a single-tier switched fabric with multiple parallel planes, and (c) a two-tier leaf-spine fabric.}
    \label{fig:topology}
\end{table*}

\subsection{Topology and Scale}
\label{sec:3.1.1}


Scale-up networks favor low-diameter topologies to limit communication latency while providing high aggregate bandwidth. Their achievable scale is jointly constrained by endpoint degree, switch radix and capacity, link bandwidth and reach, and the number of parallel fabric planes. Different topology classes shift the bottleneck among these factors: direct meshes are primarily limited by endpoint SerDes count and link growth, single-tier fabrics by switch radix and capacity, and two-tier fabrics by inter-tier bandwidth and the additional switching stage.

Here, a \emph{single-tier fabric} contains one logical switching level between endpoints, although it may use multiple parallel switches or planes. A \emph{two-tier fabric} contains two switching levels, conventionally leaf and spine; a cross-leaf path therefore traverses a source leaf, a spine, and a destination leaf. The tier count refers to the switching hierarchy rather than the number of physical switches or switch traversals.

\begin{figure}[t]
    \centering
    \includegraphics[width=\linewidth]{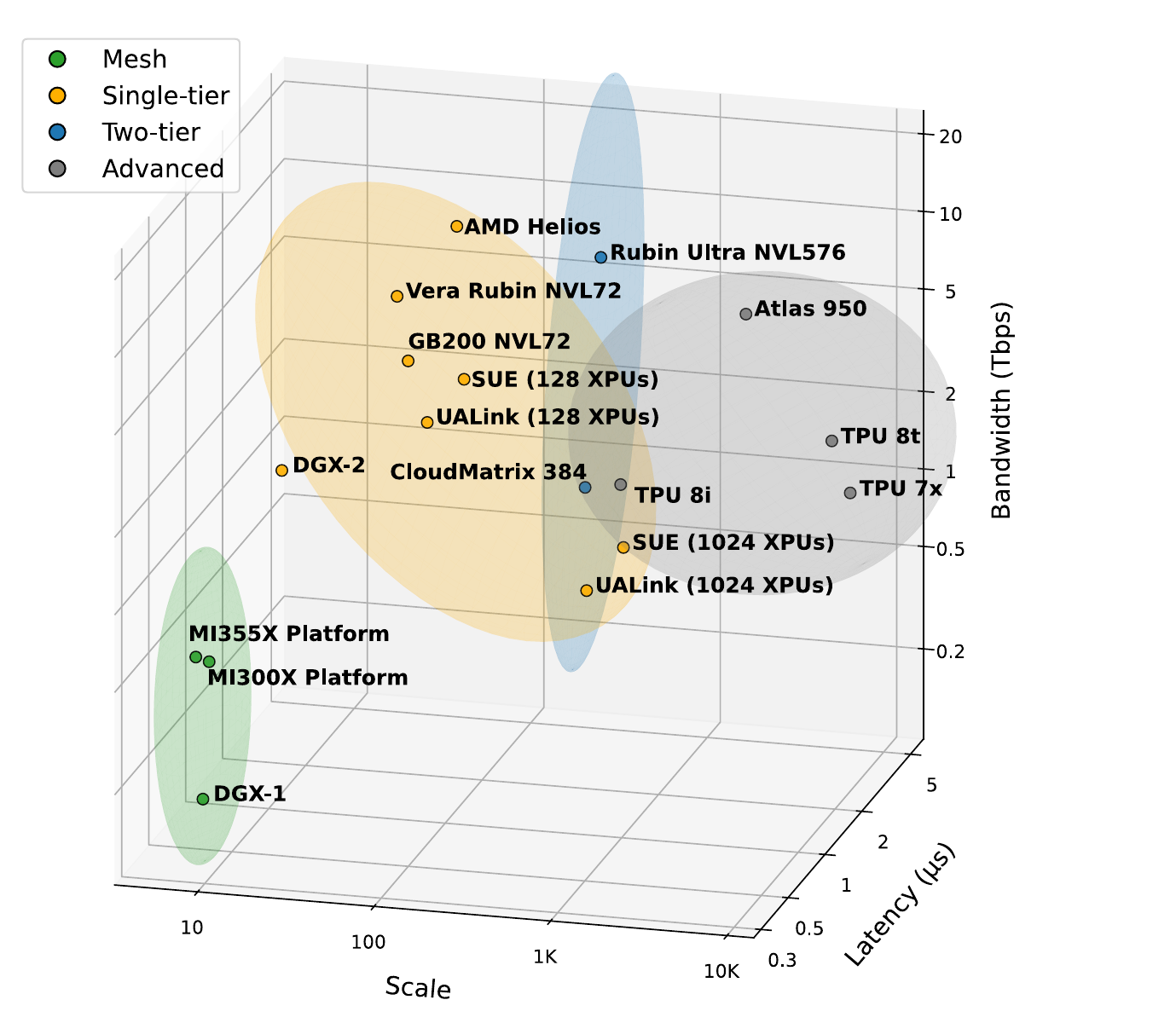}
    \caption{End-to-end latency ($\mu s$), maximal peer-to-peer bandwidth (Tbps) and scale of important scale-up superpod instances and standards against their respective topologies. The reported latency values are drawn from three types of sources: estimates based on system scale, experimental measurements, and publicly available specifications.
    }
    \label{fig:3d_graph}
\end{figure}

Figure~\ref{fig:topology} illustrates the representative structures. Table~\ref{tab:topology_properties} compares their structural properties and scaling constraints, while Table~\ref{tab:topo2} summarizes the topology support and documented scale of the surveyed fabrics. Figure~\ref{fig:3d_graph} further relates topology to the reported latency, peer-to-peer bandwidth, and system scale. These topology classes do not imply a monotonic performance ordering: increasing scale generally requires trading shorter paths and simpler control for additional switching resources and more complex traffic management.

\smalltitle{Mesh Topology.}
A full mesh provides a direct link between every pair of accelerators, eliminating intermediate switches and their associated latency and contention. However, each endpoint requires \(N-1\) connections and the total link count grows as \(O(N^2)\), making the design practical mainly at board or box scale. Moreover, the provisioned bandwidth is tied to fixed endpoint pairs; communication patterns that use only a subset of these pairs can leave a substantial fraction of the links idle.

A partial mesh reduces endpoint degree and link count by removing selected direct connections, but traffic between non-adjacent endpoints must then traverse intermediate accelerators or an external fabric. NVIDIA DGX-1, for example, connects eight GPUs using a partial cube mesh and relies on PCIe for communication between GPUs without direct NVLink connections~\cite{Nvidia_DGX_1,wang2019blinkfastgenericcollectives}. Partial meshes therefore exchange some of the full mesh's direct-connect simplicity for lower wiring cost and greater implementation flexibility.   


\smalltitle{Single-tier Topology.} A single-tier fabric connects endpoints through one logical switching level, which may consist of one switch, a multi-ASIC logical switch, or multiple parallel planes, as shown in Figure~\ref{fig:topology}(b). Parallel planes increase aggregate bandwidth and path redundancy without increasing the logical topology depth.

The scale of each plane is determined by its endpoint-facing radix and switching capacity, the number of ports assigned to each accelerator, and the target per-endpoint bandwidth. Port bifurcation increases the number of logical ports by dividing a high-rate physical port, but it does not increase total switching capacity. Logical port count therefore cannot be interpreted directly as accelerator count when an accelerator attaches through multiple ports or planes.

Compared with a direct mesh, switching enables bandwidth to be allocated more flexibly between arbitrary endpoint pairs and avoids quadratic growth in endpoint degree. The corresponding trade-off is contention at switch output ports and internal switching resources. Current specifications such as SUE, UALink, and EthLink permit single-tier domains of up to 1,024 accelerators~\cite{SUE,UALink_spec,EthLinkSpec}, although the achievable per-accelerator bandwidth depends on the switch configuration and number of fabric planes.

\smalltitle{Two-tier Topology}. A two-tier fabric uses a leaf-spine organization to extend the domain beyond the endpoint-facing radix of a single switching tier. Its achievable scale depends on both leaf and spine radix, the allocation of leaf uplinks, and the number of parallel planes. Because these systems commonly span multiple racks, they typically use optical links throughout the fabric or a hybrid design with copper for short-reach endpoint-to-leaf connections and optics between leaf and spine switches.

Publicly documented examples include Huawei CloudMatrix384, which uses a full-optical fabric~\cite{zuo2025servinglargelanguagemodels,semianalysis2025huawei}, and the disaggregated ETH+ UPN~512 architecture~\cite{ETHPlusUPN}. Compared with a single-tier fabric, the additional switching level improves reach and scale but introduces longer and potentially non-uniform paths, greater inter-tier contention, and stronger requirements for routing, congestion control, reliability, and failure handling.

\begin{table*}[t]
\centering
\caption{Protocol stack features of surveyed scale-up fabric systems.}
\label{tab:protocol}
\small
\setlength{\tabcolsep}{3pt}
\renewcommand{\arraystretch}{1.2}
\begin{tabularx}{\textwidth}{l c c >{\centering\arraybackslash}X c c}
\toprule
\multirow{2}{*}{\textbf{Solutions}} &
\multicolumn{3}{c}{\textbf{PL and DL}} &
\multicolumn{2}{c}{\textbf{NL}} \\
\cmidrule(r){2-4}\cmidrule(l){5-6}
 & LLR & CBFC & custom opt. & Routing & Fault-Tolerance \\
\midrule
NVLink & \checkmark & \checkmark & / & centralized & \checkmark \\
Infinity Fabric & \checkmark(-) & \checkmark(-) & \checkmark(-) & - & - \\
UB-2.0 & \checkmark & \checkmark & FEC \& interleave mode & centralized & \checkmark \\
SUE & \checkmark & \checkmark & FEC \& interleave mode & - & \checkmark \\
SUE-T & / & / & / & / & / \\
UALink & \checkmark & & interleave mode & centralized & \checkmark \\
ETH+ & \checkmark & \checkmark & optional FEC bypass & centralized (ANM) & path failover \\
EthLink & \checkmark & \checkmark & FEC \& interleave mode & - & \checkmark \\
ICI & \checkmark & \checkmark & - & centralized & \checkmark \\
TTP/TTPoE & - & - & - & pre-configured & - \\
ESUN & \checkmark & \checkmark & FEC \& interleave mode & / & / \\
\bottomrule
\end{tabularx}

\vspace{0.5em}

\begin{tabular*}{\textwidth}{@{\extracolsep{\fill}}l *{9}{c}@{}}
\toprule
\multirow{2}{*}{\textbf{Solutions}} & \multicolumn{4}{c}{\textbf{Transport Layer}} &
\multicolumn{4}{c}{\textbf{Load Balancing}} & \textbf{Header/Packing} \\
\cmidrule(lr){2-5}\cmidrule(lr){6-9}
 & Reliability & CC & FC & Security & Transport & NL & Per-flow & Per-packet & \textbf{Optimizations} \\
\midrule
NVLink & [\checkmark] & - & - & \checkmark & - & - & - & - & \checkmark \\
Infinity Fabric & - & - & - & - & - & - & - & - & - \\
UB-2.0 & [\checkmark] & [\checkmark] &  & [\checkmark] & [\checkmark] & \checkmark & \checkmark & \checkmark & \checkmark \\
SUE & [\checkmark] & [\checkmark] &  & [\checkmark] & [\checkmark] &  & [\checkmark] & [\checkmark] & \checkmark \\
SUE-T & [\checkmark] & [\checkmark] & / & \checkmark] & [\checkmark] & / & [\checkmark] & [\checkmark] & \checkmark \\
UALink & \checkmark &  & \checkmark & [\checkmark] & \checkmark & - & - & - & \checkmark \\
\new{ETH+} & \new{[\checkmark]} & \new{[\checkmark]} & \new{-} & \new{-} & [\checkmark] & [\checkmark] & [\checkmark] & [\checkmark] & \new{\checkmark} \\
EthLink & [\checkmark] & [\checkmark] &  & [\checkmark] & - & - & [\checkmark] & [\checkmark] & \checkmark \\
ICI & - & - & - & - & - & \checkmark & / & / & \checkmark \\
TTP/TTPoE & \checkmark & \checkmark & \checkmark & - & - & - & - & - & \checkmark \\
ESUN & / & / & / & / & / & \checkmark & - & - & \checkmark \\
\bottomrule
\end{tabular*}

\vspace{2pt}

\begin{minipage}{0.96\linewidth}

\raggedright
\tiny
$^{*}$ UALink provides end-to-end error detection but no retransmission.

\small
\centering
$\checkmark$: feature is present; $[\checkmark]$: feature is optional; $/$: not applicable; $-$: undetermined for lack of related public information.

\end{minipage}

\end{table*}

\subsection{Communication and Transport}

Communication protocols used for scale-up networks generally follow a lightweight design to ensure minimal processing overhead and decreased software latency. The following discussion features a bottom-up analysis of scale-up protocols against a modified five-layer TCP/IP model, excluding the topmost application layer and including an additional transaction layer. \fix{We note that the introduction of the transaction layer can be a major source of confusion for researchers focusing on conventional networking protocol stacks, so a detailed discussion of its role is provided later. We examine the header and data-packing optimizations that span multiple layers at the end of this subsection. To keep the discussion concise, we illustrate each architectural characteristic using only a few representative examples, while details concerning all eleven surveyed protocols will be presented in respective tables. }

\subsubsection{Physical and Data Layer}
\label{sec:3.2.1}

The physical (PL) and data link (DL) layers define how data is encoded as bits, framed, and transmitted. Scale-up protocols employ extensive optimizations on these two layers to push the latency and goodput performance to theoretical limits.

LLR and CBFC are two advanced L2 features supported by the majority of scale-up standards. LLR provides hop-by-hop reliability by locally retransmitting corrupted frames between adjacent hops. This improves the latency and goodput performance by reducing the number of expensive end-to-end replays; CBFC is a proactive flow control mechanism: the receiver periodically issues credits to the sender based on its buffer availability, and each transmission consumes a number of credits proportional to the payload size. CBFC offers several advantages over the standard PFC (Priority Flow Control) in RoCE networks: reduced head-of-line (HoL) blocking, higher buffer utilization and finer traffic-class granularity \cite{UEC2026Spec}. These two mechanisms are widely co-adopted by both mature fabrics (NVLink, ICI) and new Ethernet-derived standards (SUE, UALink, UB-2.0, ESUN) to form a highly efficient lossless environment. 

The use of a flat L2 addressing scheme is another common feature shared by the majority of scale-up protocols. Because scale-up protocols primarily target operation within a single domain where the number of accelerator endpoints can be pre-determined, the conventional two-tier IP/MAC structure gives way to a more natural L2-only addressing scheme. For example, NVLink bypasses the standard IP+MAC addressing scheme using Local Identifiers (LIDs), UB-2.0 uses a 32-bit Concise Network Address (CNA) for addressing within a single domain, UALink relies on a flat 10-bit accelerator identifier within a single domain, and ETH+ forwards on a 4-byte Unified Identifier (UID) carried in an IPv6-like base header in place of MAC addressing~\cite{ETHPlusSpec,ETHPlusFraming}. In the rare case where cross-domain operations are defined for a scale-up protocol, the addressing simply uses a lengthened local identifier instead of reverting back to a two-tier structure. For example, UB-2.0 uses the 128-bit Network Address (NA) in place of CNA for cross-domain communications. 
 

Aside from the aforementioned common mechanisms, several standards introduce PHY-specific optimizations to reduce link latency. For instance, SUE~\cite{SUE} and EthLink~\cite{EthLinkSpec} support reduced-overhead FEC, such as RS(272), at 400/800~Gb/s link rates, trading coding gain for lower latency. UB~2.0~\cite{UBSpec}, by contrast, is not derived from Ethernet: it defines its own link coding and RS(128,120) FEC while also providing an IEEE~802.3-compliant electrical mode. These designs illustrate that scale-up fabrics may reduce coding and interleaving overhead, or define a dedicated link layer, to meet stringent latency targets.

\subsubsection{Network Layer}

The network layer of a scale-up protocol differs from that of a conventional networking protocol in several significant ways. In terms of routing, most existing scale-up fabrics use a centralized manager to handle address assignment, topology discovery and routing configuration. These procedures are typically performed during the fabric initialization, freeing switches from running individual copies of routing instances. For example, NVLink, UB-2.0, UALink and ETH+ all use one or more centralized agents (Fabric Manager and NVLink Subnet Manager for NVLink; Unified Bus Fabric Manager and Unified Bus Controller(s) for UB; Pod Controller assisted by Switch Management Agents and Node Management Agents for UALink; Auto Network Management (ANM) system for ETH+) per domain for address assigning, topology sensing and forwarding table programming. Similarly, Google's TPU cluster deploys a centralized \verb|libtpunet| library instance within the user's job for address assignment and routing configuration \cite{TPUv4}. It is worth noting that these centralized agents generally support other functionalities of failure detection, health monitoring and telemetry. The centralized, SDN-style routing scheme in question is best understood as one of the many features of the underlying centralized management infrastructure.

In addition, the network layer should provide failure recovery mechanisms to ensure minimal performance degradation in case of failure. For single-tier fabrics, complex fault-tolerant routing is unnecessary because the system can simply turn off the failed links/ports and route through functional ones. For example, SUE supports automatic migration to other functional ports for sprayed traffic that encounters a failed port. For Mesh networks, rerouting through peer accelerator or redundant paths is necessary in case of a link failure (as discussed in \S~\ref{sec:3.1.1}). As an example, Huawei designed the All-Path Routing (ARP) for its UB-Mesh, which takes detours through peer NPUs around the failed link \cite{liao2025ubmesh}. Torus, two-tier and more complex networks require advanced fault-tolerant routing to decide which of the multiple paths between the sender and receiver should be used. Google's \verb|libtpunet| employs wild-first routing, which allows taking at most one wild hop in each of the x, y, z dimensions around the link/port failure, before restoring shortest-path-first routing \cite{TPUv4}. To sum up, single-tier networks allow for simpler fault-handling logic at the network layer, which again explains its popularity within industry realizations. Other topologies require more complex failure-aware rerouting schemes that incur additional overhead at the accelerators or switches.

\begin{figure*}[t]
    \centering

    \subfloat[RoCEv2 packing.\label{fig:hdr-ethernet}]{%
        \includegraphics[
            width=0.5\textwidth,
            trim=87pt 292pt 49pt 30pt,
            clip
        ]{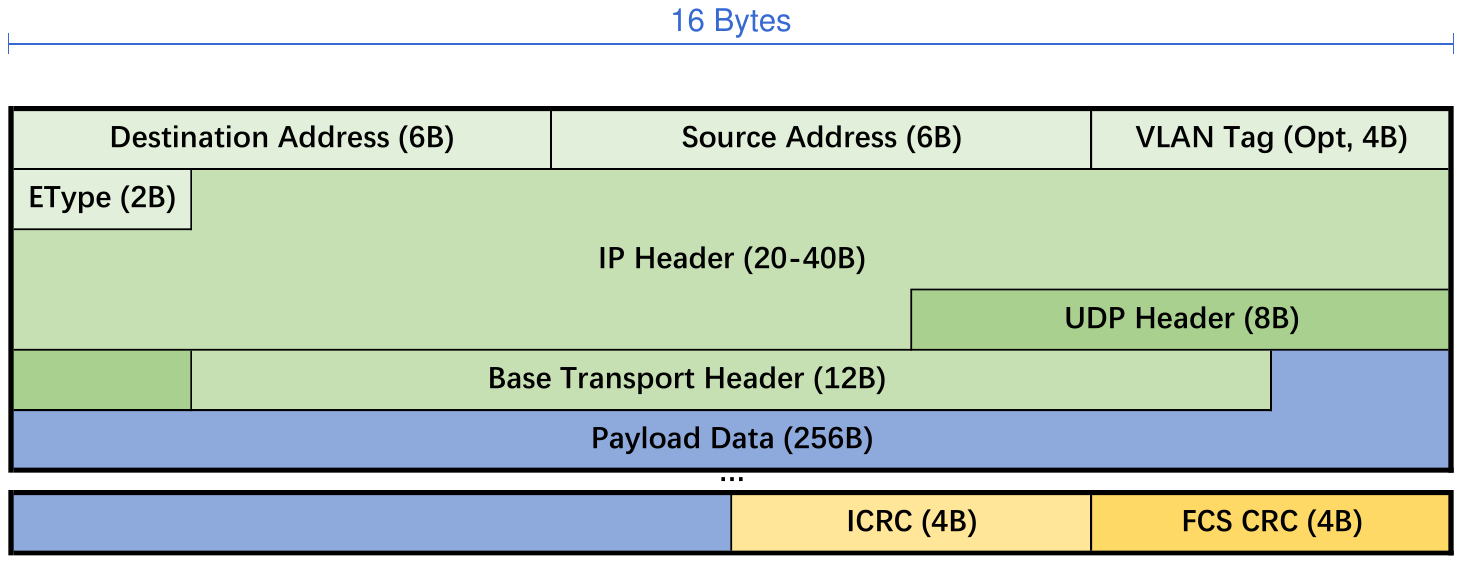}%
    }

    \vspace{0.2em}
    
    \subfloat[SUE packing (AFH Gen2 in red).\label{fig:hdr-sue}]{%
        \includegraphics[
            width=0.5\textwidth,
            trim=87pt 412pt 49pt 25pt,
            clip
        ]{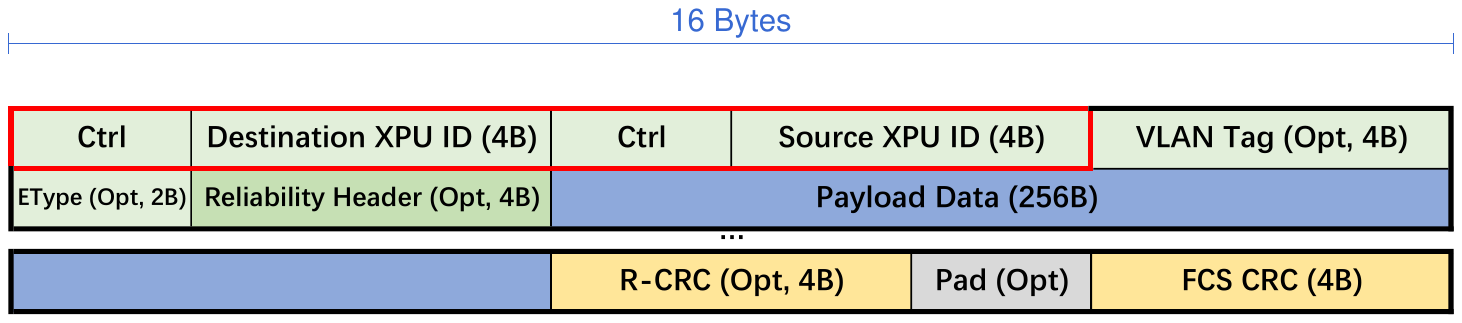}%
    }
    \hfill
    \subfloat[EthLink packing (OEFH in red).\label{fig:hdr-ethlink}]{%
        \includegraphics[
            width=0.5\textwidth,
            trim=75pt 373pt 53pt 30pt,
            clip
        ]{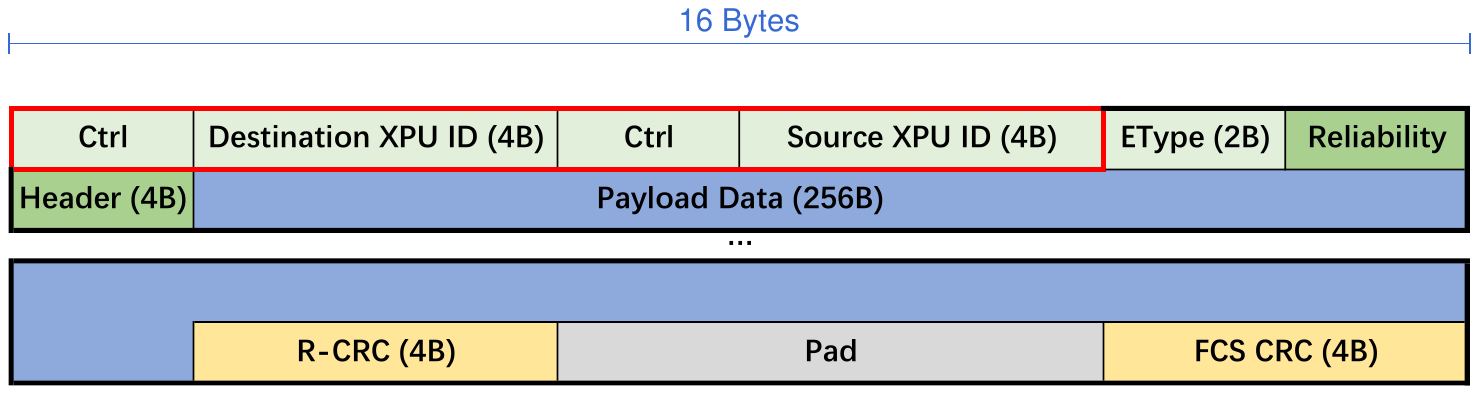}%
    }

    \vspace{0.2em}

    \subfloat[ETH+ identifier-based-mode packing (UID base header in red).\label{fig:hdr-ethplus}]{%
        \includegraphics[
            width=0.5\textwidth,
            trim = 122pt 277pt 41pt 21pt,
            clip
        ]{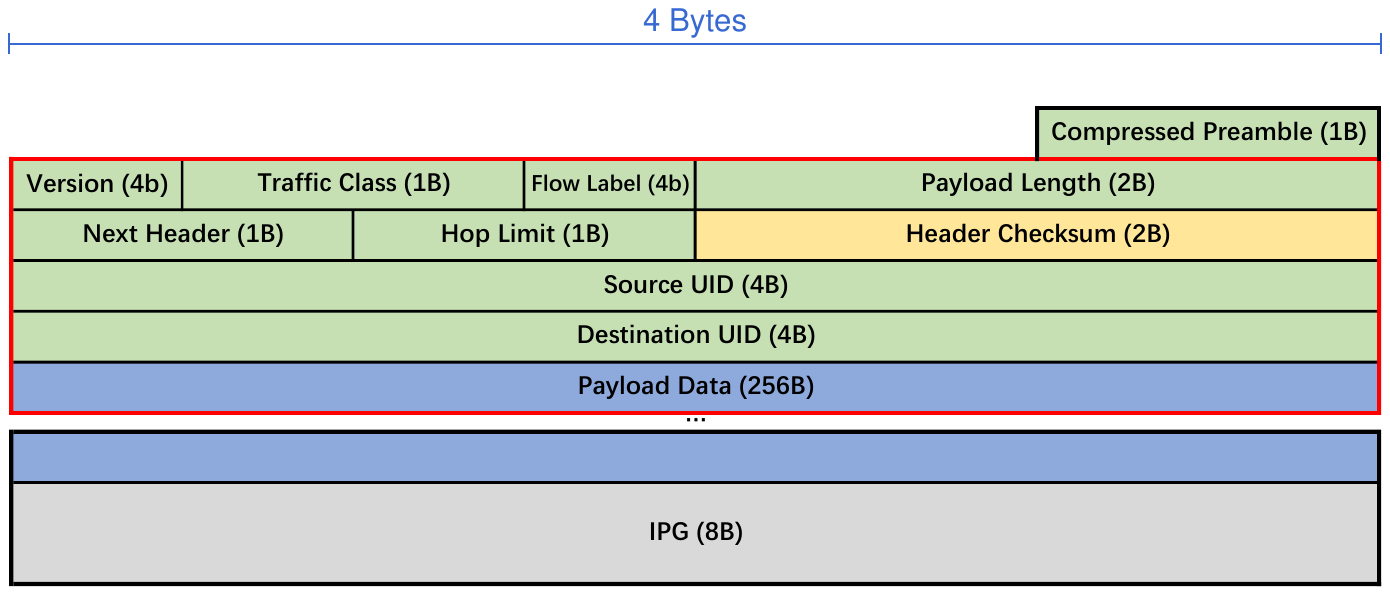}%
    }
    \hfill
    \subfloat[ESUN packing.\label{fig:hdr-esun}]{%
        \includegraphics[
            width=0.5\textwidth,
            trim=75pt 388pt 53pt 18pt,
            clip
        ]{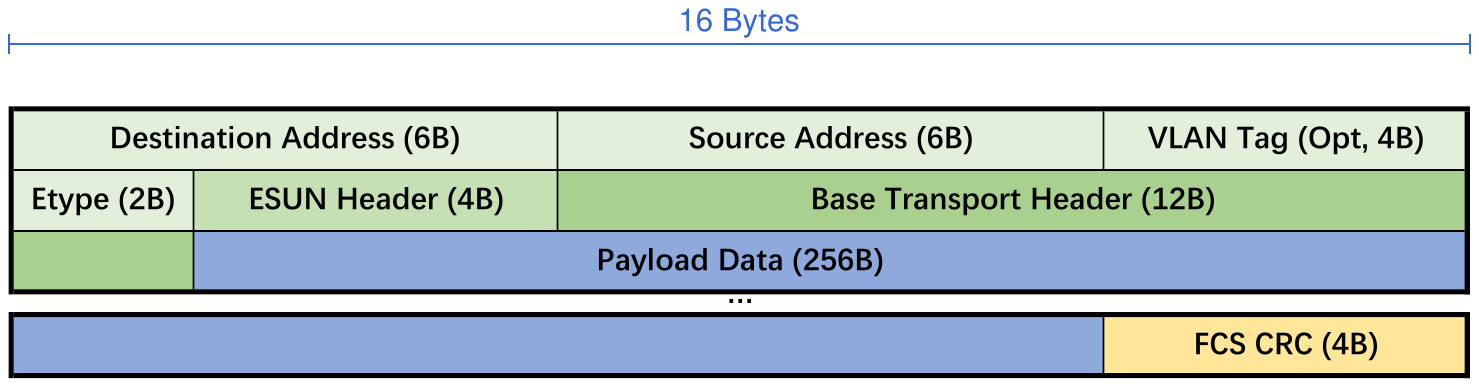}%
    }

    \vspace{0.2em}

    \subfloat[UALink packing.\label{fig:hdr-ualink}]{%
        \includegraphics[
            width=0.5\textwidth,
            trim=30pt 160pt 133pt 15pt,
            clip
        ]{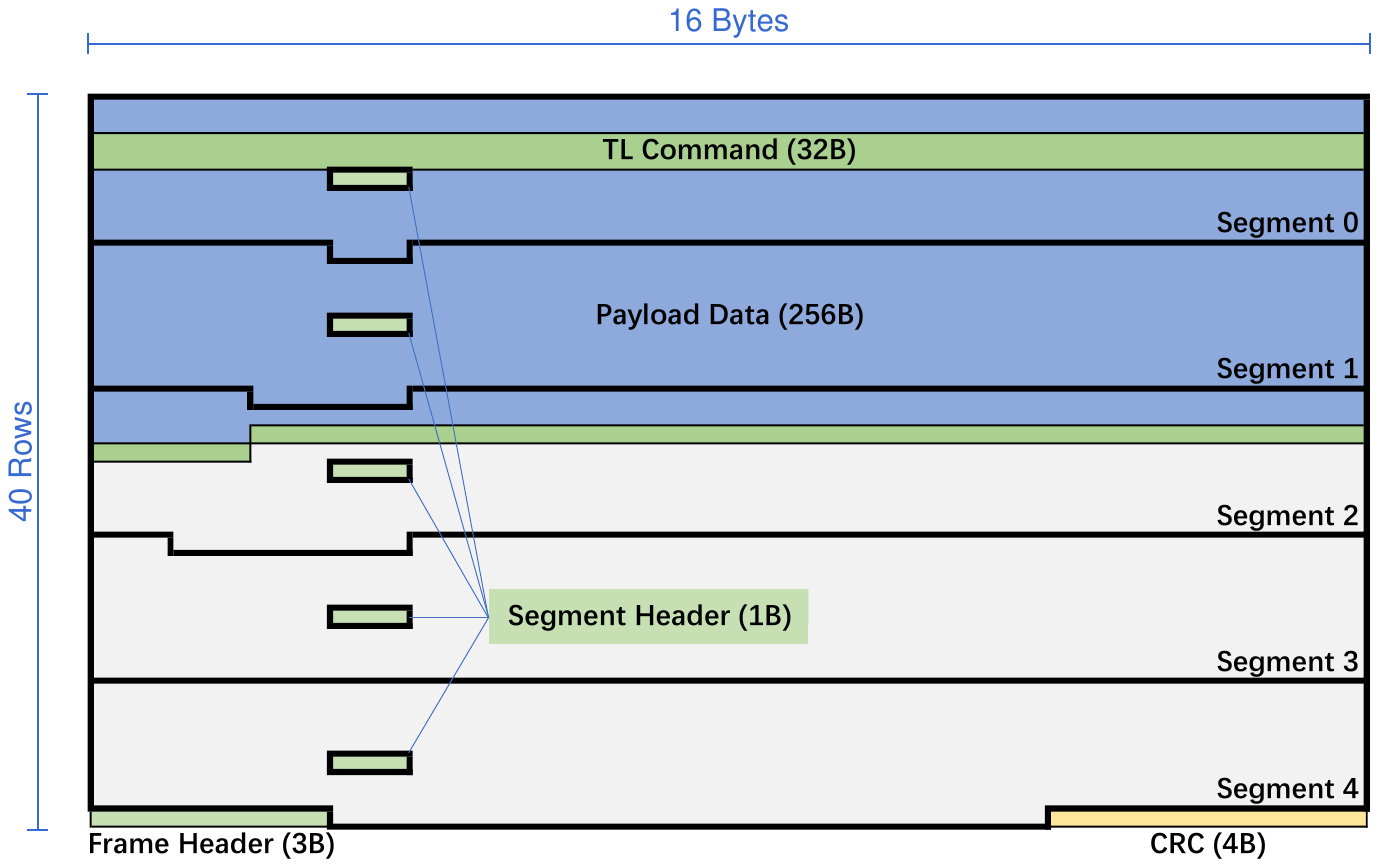}%
    }
    \hfill
    \subfloat[NVLink packing.\label{fig:hdr-nvlink}]{%
        \includegraphics[
            width=0.42\textwidth,
            trim=58pt 145pt 139pt 27pt,
            clip
        ]{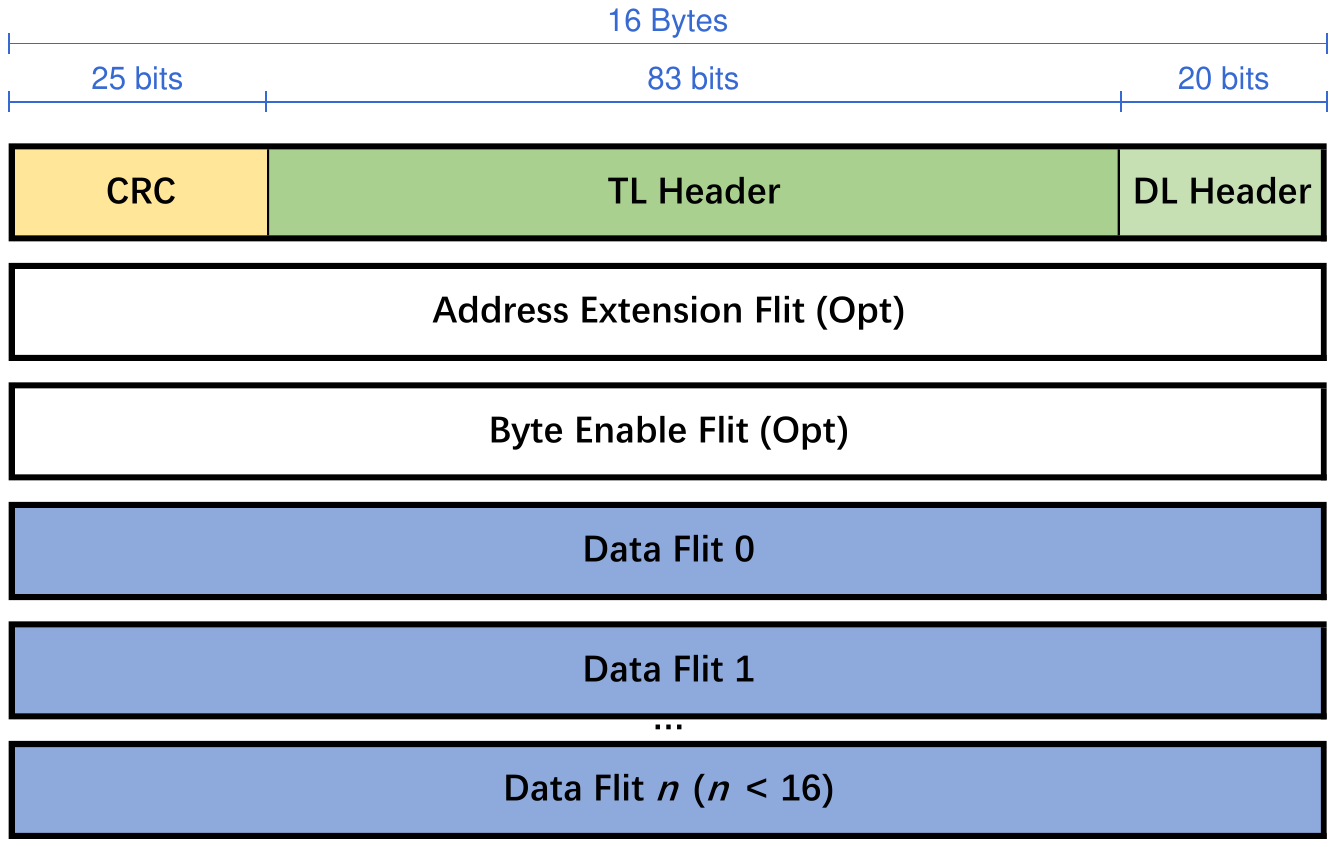}%
    }

    \caption{Data packing formats of (a) RoCEv2 and (b)–(g) representative scale-up fabric systems. Payload fields are
    vertically compressed for readability.}
    \label{fig:header-comparison}
\end{figure*}

\subsubsection{Transport Layer}
\label{sec:3.2.3}

The transport layer of a conventional networking protocol provides end-to-end functions, including reliability, flow control, congestion control, and ordering guarantees. These functions remain relevant in scale-up fabrics, but the strength of the required guarantees and the placement of the corresponding mechanisms in the protocol stack vary across systems. In particular, ordering requirements depend on the semantics of the operations and the completion model. RDMA reads, writes, and atomic operations in reliable connection mode require strict ordering within each queue pair, whereas weakly ordered memory-semantic transactions may be delivered out of order as long as dependencies expressed through fences or barriers are preserved. Accordingly, protocols such as UB-2.0, SUE, UALink, EthLink, and ETH+ define both ordered and relaxed/unordered operating modes.

Similarly, the placement of reliability and traffic-control functions depends on both the topology and the required service contract. In a shallow mesh or single-tier fabric, congestion and failures remain sufficiently localized that some end-to-end functions can be simplified or delegated to the link and network layers. Link-level retransmission and credit-based flow control, for example, can provide efficient local recovery and loss prevention without requiring equivalent mechanisms at higher layers. SUE and SUE-T illustrate this principle. SUE offers a lightweight variant that omits a distinct transport layer and relies primarily on lower-layer mechanisms. SUE-T states that congestion control is unnecessary in single-tier networks and permits Go-Back-N retransmission to be disabled in selected configurations. These design choices can reduce software overhead and communication latency at the cost of reduced compatibility with more complex networks.

As the number of switching stages and the degree of path diversity increase, congestion and failures are more likely to have nonlocal effects. The need for end-to-end congestion control and reliability mechanisms therefore increases. For example, SUE-T identifies congestion control as necessary in three-tier networks and describes how it can be integrated. TTPoE has been used in two-tier fabrics and provides window-based flow control, end-to-end retransmission, and congestion management \cite{Quinnell2024ttpoe}. Scale-up protocol families that support arbitrary topologies further illustrate this trade-off. UB-2.0 defines two transport modes, one reliable and ordered and the other unreliable and unordered. It gives users considerable flexibility in selecting and configuring mechanisms for congestion control, load balancing, and retransmission. ETH+ also supports arbitrary topologies. Its transport layer implements congestion control, flow control, load balancing, and reordering as optional modules that can be enabled or disabled according to system requirements.

Thus, topology shapes transport-layer design by influencing the scope and complexity of the required communication mechanisms. A simple topology can allow some transport functions to be bypassed or simplified, reducing software overhead and resource consumption at the accelerator endpoint. This helps explain the popularity of single-tier topologies in medium-scale scale-up fabrics. In systems with more complex topologies and traffic patterns, a feature-rich transport layer and its associated software overhead are harder to avoid.

\subsubsection{Transaction Layer} The primary functionality of a transaction layer is to translate between the lower-level bit/byte-stream semantics of a packet switching fabric and a higher-level semantics required by upper-layer software. For example, PCIe operates on the basis of memory commands between the CPU and peripheral devices instead of raw data passing. As such, PCIe defines a transaction layer that converts upper-layer commands/transactions into packets to be carried by the underlying interconnect networks. Similarly, iWARP defines several middle layers sitting between upper-layer protocols and the standard TCP/IP stack; specifically, the RDMAP layer  provides the read/write/atomic semantics required by RDMA, and is therefore functionally equivalent to a transaction layer \cite{RFC5040, mittal2018IRN}. Within scale-up networks, accelerators communicate with peers through memory-semantic load/store operations rather than direct data passing. This semantics mismatch again dictates a transaction layer in the scale-up protocol stack.

Conceptually, the transaction layer acts as an interface to upper-layer software and manages the states for each transaction before it completes. For our discussion, the most important characteristic of a scale-up transaction layer is its support for the \textbf{memory semantics} operations, which are discussed in \S\ref{sec:3.3}

\subsubsection{Load Balancing}
\label{sec:load_balancing}

Load balancing distributes traffic across multiple network paths between the source and destination to improve bandwidth utilization and reduce contention. Load balancing in scale-up fabrics differs primarily along two independent dimensions: placement and granularity. For placement, load balancing decisions can be made at endpoints by the transport layer or at intermediate switches by the network layer. For granularity, load balancing can operate on a coarser-grained, per-flow basis (similar to ECMP forwarding), and on a fine-grained, per-packet basis (similar to packet spraying). Table~\ref{tab:protocol} reports load balancing supported by each scale-up fabric family across these two dimensions. The following paragraphs examine the strengths and weaknesses of each load balancing implementation and how they correlate with other system attributes including topology, ordering requirements, etc.

Generally speaking, the fabric topology has a first-order effect on where load balancing is required. In a fully connected mesh fabric with no forwarding at accelerators, there exists exactly one direct link per accelerator pair, making load balancing unnecessary. In typical single-tier, multi-rail networks, each accelerator endpoint must select which rail to send traffic to, after which the rail switch can simply forward the traffic out of the only link to the destination accelerator. As such, single-tier, multi-rail networks require only accelerator-side, transport-layer load balancing. For example, SUE and SUE-T permit load balancing to be implemented at the accelerator endpoint through either custom software logic or a hardware block. UALink and EthLink similarly permit accelerators to distribute traffic among multiple ports and protocol instances. Two-tier and more complex networks additionally require switch-side, network-layer load balancing to properly distribute traffic across inter-switch links. For instance, UB-2.0 explicitly defines per-packet and per-flow load balancing at the network layer and per-channel load balancing at the transport layer. ETH+ allows accelerators to use an optional transport scheduler to redirect traffic around congested or failed links, while ETH+ switches can select among eligible egress ports based on the flow-label field. ICI employs a static load balancing algorithm at the network layer: the routing software programs exactly one path per source\hyp{}destination pair at job startup, achieving load balancing through coordinated route assignments.  

In terms of granularity, the coarser-grained per-flow load balancing distributes traffic less evenly than the fine-grained per-packet load balancing, particularly in multi-tier networks, where hash polarization may occur. On the other hand, per-packet load balancing is much more challenging to implement, incurs greater software processing overhead and causes packet reordering. While reordering is less of a problem in the scale-up setting due to the weakly-ordered nature of memory-semantic operations, the excessive processing needed for per-packet load balancing makes it better suited to switches than to endpoints, consistent with the general scale-up philosophy of keeping the endpoint lightweight and offloading functions to the network. Nevertheless, the distinction between per-flow and per-packet load balancing and their mapping to accelerator endpoints and switches remain unclear due to the limited level of integration into existing scale-up families and lack of public information. SUE, SUE-T and UALink allow accelerator endpoints to perform load balancing operations without specifying whether it is per-flow or per-packet. EthLink mentions support for both per-flow and per-packet load balancing, but does not prescribe their placement. Exceptions include UB-2.0, which supports per-flow and per-packet load balancing at both the transport and network layers, and ETH+, which supports per-packet and per-flow load balancing at the network layer.

    








    

\begin{table*}[t]
\centering
\small
\setlength{\tabcolsep}{4pt}
\caption{Normalized encoding efficiency for a 256-B data-bearing transfer. Exact values require a fully specified packet format. Upper bounds include only publicly documented mandatory overhead and may therefore overestimate the achievable efficiency.}
\label{tab:normalized_encoding_efficiency}
\begin{tabular}{
    p{0.105\textwidth}
    p{0.145\textwidth}
    p{0.105\textwidth}
    p{0.13\textwidth}
    p{0.445\textwidth}}
\toprule
\textbf{Fabric} &
\textbf{Counted wire bytes, \(D_{\mathrm{wire}}\)} &
\textbf{Efficiency, \(\eta_{\mathrm{enc}}\)} &
\textbf{Result type} &
\textbf{Accounting boundary and limitation} \\
\midrule

RoCEv2 &
334 B / 354 B &
76.6\% / 72.3\% &
Exact &
IPv4/IPv6 RDMA Write Only packet without VLAN. The accounting includes the Ethernet header, IP, UDP, BTH, RETH, ICRC, and FCS. Preamble and inter-packet gap are excluded. \\

Ultra Ethernet &
At least 360 B &
At most 71.1\% &
Upper bound &
Calculated from the publicly reported 104-B header of a basic UET packet. Optional headers, trailers, and physical-layer overhead can further reduce the efficiency. \\

SUE &
At least 278--284 B &
90.1\%--92.1\% &
Upper-bound range &
Includes a 6- or 12-B AFH Gen2 header, an 8-B reliability header, a 4-B reliability CRC, and a 4-B FCS. Packed transaction-control fields and physical-layer overhead are excluded. \\

SUE-T + ESUN &
At least 302 B &
At most 84.8\% &
Conditional upper bound &
Single 256-B write, CRC enabled: 14-B Ethernet + 4-B ESUN + the SUE-T document's stated 8-B RODL + 12-B uncompressed semantic header + 4-B UET CRC + 4-B FCS. No VLAN, XPU extension, or security header.  \\

UALink (200G)&
640 B &
40.0\% &
Isolated-message scenario &
An isolated transfer carrying 256 B of useful data occupies one 640-B DL flit; the remaining 384 B comprise non-payload bytes, including metadata and padding. The efficiency can increase when multiple transactions are packed into the same flit. \\

ESUN &
At least 278 B &
At most 92.1\% &
Upper bound &
Includes a 14-B Ethernet header, a 4-B ESUN header, and a 4-B FCS. ESUN specifies network functions only through layer 3; additional higher-layer header bytes are therefore excluded. \\

EthLink &
At least 266 B &
At most 96.2\% &
Upper bound &
Includes the publicly documented 6-B optimized Ethernet header and a 4-B FCS. Additional transaction, integrity, alignment, and
physical-layer fields are not included. \\

\bottomrule
\end{tabular}
\end{table*}

\subsubsection{Header and Packing Efficiency}
\label{sec:header_efficiency}

Scale-up protocols reduce serialization overhead by using compact, specialized packet formats. Figure~\ref{fig:header-comparison} illustrates the publicly documented formats of representative protocols. Unlike RoCEv2, which encapsulates an RDMA operation through Ethernet, IP, UDP, and RDMA headers, scale-up protocols commonly remove intermediate network layers, compress addressing fields, or combine reliability and transaction metadata into fabric-specific headers. These optimizations are particularly important for small transfers, for which fixed protocol overhead accounts for a substantial fraction of the bytes transmitted on the wire.

A direct comparison is nevertheless difficult because public specifications expose different protocol boundaries. Some describe complete serialized frames, whereas others disclose only selected headers or fixed-size link units. We therefore normalize the comparison to a 256-B data-bearing transfer and define its encoding efficiency as $\eta_{\mathrm{enc}} = \frac{D_{\mathrm{payload}}}{D_{\mathrm{wire}}}$, where \(D_{\mathrm{payload}}\) is the useful data carried by the
transfer and \(D_{\mathrm{wire}}\) includes the payload together with the mandatory protocol overhead and padding counted within the specified boundary. Unless otherwise stated, physical-layer coding, FEC, preamble, inter-packet gap, retransmissions, and idle cycles are excluded. When the public documentation does not specify a complete packet format, only the documented mandatory fields are counted, and the resulting efficiency is reported as an upper bound.

Table~\ref{tab:normalized_encoding_efficiency} shows that the disclosed scale-up formats generally reduce header overhead relative to the multi-layer RoCEv2 stack. However, the numerical results are not all equally definitive. RoCEv2 can be calculated from a complete encapsulation format, whereas the values for Ultra Ethernet, SUE, ESUN, SUE-T + ESUN and EthLink are upper bounds because some transaction, alignment, integrity, or optional fields are not fully specified in public documents. For ETH+ and current NVLink generations, the available information is insufficient to reconstruct a complete serialized packet, so no normalized value is reported.

Fixed-size transfer units also require separate interpretation. For example, an isolated UALink 200G transfer carrying 256 B of useful data within one 640-B DL flit has an encoding efficiency of 40\% at the DL-flit boundary. This value does not represent steady-state link utilization when multiple messages or protocol data units share the remaining flit capacity. More generally, the normalized values in Table~\ref{tab:normalized_encoding_efficiency} characterize only encoding and framing overhead. They do not include flow-control stalls, retransmissions, topology-dependent traffic, or implementation-specific physical-layer overhead. The comparison therefore demonstrates the structural advantage of compact scale-up packet formats without implying that header efficiency alone determines application-level throughput.

\smalltitle{Summary:}
Table~\ref{tab:protocol} summarizes the protocol-stack features of the scale-up fabrics examined in this paper. Viewed from the bottom up, their protocol designs follow a consistent performance-oriented principle. The physical and data-link layers provide efficient local loss recovery and flow control, while the network and transport layers are simplified when topology-specific assumptions permit. Compact packet formats and specialized transaction semantics further shorten the communication critical path. Scale-up protocol stacks are therefore not merely faster versions of conventional scale-out stacks; they are structurally specialized for a more tightly controlled environment and more aggressive bandwidth and latency targets.

\begin{table*}[t]
\centering
\caption{Fabric primitives and accelerator-visible realizations of the surveyed scale-up systems. Missing public information is not interpreted as evidence that an implementation lacks a capability.}
\label{tab:semantics}
\small
\setlength{\tabcolsep}{4pt}
\renewcommand{\arraystretch}{1.15}
\begin{tabular}{
    p{0.15\linewidth}
    p{0.30\linewidth}
    p{0.47\linewidth}}
\toprule
\textbf{System} &
\textbf{Fabric Primitive} &
\textbf{Representative Accelerator-Visible Realization} \\
\midrule

NVLink/NVSwitch
~\cite{NVLink,nvshmem_docs,NCCLDeviceAPI} &
Peer-memory transactions, atomics, multicast, and reduction support, depending on generation &
CUDA peer-memory access, NVSHMEM put/get and atomics, and NCCL collectives or Device API operations; selected operations can be initiated directly by GPU code. \\

Infinity Fabric~\cite{InfinityFabric,rocSHMEM} &
Peer-memory and product-dependent coherent or non-coherent transactions &
AMD runtime data movement, rocSHMEM device-side put/get and tile operations, and collective libraries; the exposed contract depends on the CPU, GPU, or APU generation. \\

ICI~\cite{Jouppi2023TPUv4,TPUv4} &
Point-to-point and collective communication &
Communication is generated by the TPU compiler and runtime; public descriptions do not expose a general accelerator-visible remote-memory
interface. \\

Unified Bus 2.0
~\cite{UBSpec,openeuler2025unifiedbus} &
DMA-style transfer and synchronization transactions &
TMA/STMA-style data movement and runtime collectives executed by device communication engines; public material does not establish ordinary remote load/store or put/get instructions. \\

SUE~\cite{SUE,SUE_Whitepaper} &
XPU-facing read, write, atomic, and explicit data-transfer transactions &
The XPU interface maps upper-layer requests onto SUE transactions; their mapping to accelerator instructions, runtimes, or DMA engines is
implementation-specific. \\

SUE-T~\cite{SUE_T_Base_1_0} &
UEC Unified Local read, write, atomic, and fetching-atomic transactions &
AXI or signal interfaces map accelerator requests to standardized semantic headers, with optional opaque accelerator extensions. Delivery/Operational ACK and memory visibility depend on the selected completion contract; ISA/runtime realization remains implementation-specific. \\

UALink~\cite{UALink_Common_2_0, UALink_spec,UALink} &
Memory-oriented read, write, atomic, and synchronization transactions; optional multicast and ReadReduce primitives, and switch-executed Broadcast, Reduce, and AllReduce block collectives &
UALink defines endpoint transaction, ordering, and completion behavior but leaves the accelerator ISA and runtime interface to individual
implementations. \\

EthLink~\cite{EthLinkSpec} &
Remote-memory and explicit data-movement transactions &
Available public material does not consistently disclose the mapping between accelerator operations, communication engines, and EthLink
transactions. \\

TTPoE~\cite{Quinnell2024ttpoe,teslamotors2024ttpoe} &
Reliable transport for upper-layer messages and data movement &
TTPoE defines transport behavior rather than one mandatory accelerator-visible operation model; semantics are supplied by the upper-layer runtime or communication library. \\

ESUN~\cite{OCP_ESUN_Launch,OCPESUN2026} &
Lower-layer Ethernet transport mechanisms &
The current specification does not prescribe accelerator-visible operations and leaves semantic implementation to upper layers. \\

T-Head ICN~\cite{THeadICN} &
Peer-device and shared-memory access &
Vertically integrated with T-Head accelerators, ICN Switch, and the associated runtime and collective library; the ISA-level mapping and
memory-ordering contract are not publicly disclosed. \\

ETH+~\cite{ETHPlusSpec,ETHPlusUPN} &
RDMA and in-network-computing substrate &
The UPN512 profile defines Load/Store, Send/Recv, and Tensor Push/Pull. These names describe accelerator-visible interfaces; they do not by
themselves establish ordinary ISA-native load/store execution. \\

\bottomrule
\end{tabular}
\end{table*}

\subsection{Communication Semantics and Executions}
\label{sec:3.3}

The transaction layer discussed in the preceding section describes how an upper-layer operation is represented as fabric requests and responses. Communication semantics instead define the contract visible to an accelerator and its software stack: which operations can be issued, who can issue them, what data objects they target, and when their effects are considered complete and visible.

These properties should be separated into three dimensions. \emph{Operation semantics} specify what is performed, such as load/store, put/get, send/receive, or a collective operation. The \emph{initiation model} specifies whether communication is issued by a host processor, a device runtime, or accelerator code. The \emph{execution path} specifies whether the operation is executed by processor instructions, a DMA engine, a network endpoint, a software proxy, or a combination of them. Similar operations may therefore have substantially different performance and ordering behavior.

\smalltitle{Semantic contract.}
Load/store operations move data between processor registers and local or remote memory. A load usually creates a data dependency because an instruction that consumes the loaded value cannot proceed until the data returns. A store, in contrast, may complete at the issuing processor before the written value becomes visible to a remote execution context. The ordering and visibility of these operations depend on the memory consistency model, the scope over which the guarantee applies, and explicit synchronization operations such as fences, barriers, or atomics. Put/get interfaces expose one-sided memory-to-memory transfers in which the initiator specifies the remote memory object. Message interfaces expose send/receive operations that typically require
matching communication at the two endpoints. Collective interfaces define group operations, such as AllReduce, AllGather, or All-to-All, whose implementation may internally use message transfers, put/get, DMA, or load/store. These classes are not mutually exclusive: a platform may use load/store for fine-grained state, put/get for explicit data movement, and collectives for large tensors~\cite{taranov2022efficientrdmacommunicationprotocols,nvshmem_docs}.

For all these interfaces, three events should be distinguished. \emph{Local completion} allows the source to reuse its local resources; \emph{remote completion} indicates that the destination interface or memory system has accepted the operation; and \emph{remote visibility} indicates that a destination execution context can observe its effect. Similarly, ordering may be scoped to an initiator, queue, stream, destination, or explicit fence rather than provided globally.

\smalltitle{Host-initiated data movement.}
A conventional DMA or RDMA path separates data movement from communication control. Technologies such as GPUDirect RDMA allow a NIC to transfer data directly between GPU memory and the network without using host memory as an intermediate buffer. However, a direct data path does not necessarily imply device-initiated control: the CPU or host runtime may still create work queues, submit descriptors, and process completion events~\cite{GPUDirectRDMA, taranov2022efficientrdmacommunicationprotocols}. This model is effective for large, planned transfers, but host interaction can become expensive when communication decisions are made inside a GPU kernel or when many small operations are interleaved with computation. Such cases motivate accelerator-initiated communication.

\smalltitle{Device-side one-sided communication.}
NVSHMEM provides a concrete implementation of device-side one-sided communication. CUDA kernel threads can issue put/get, atomic, signal, and synchronization operations on symmetric memory without returning to the CPU for each operation. The interface uses weak ordering by default: \texttt{fence} establishes scoped delivery ordering, whereas \texttt{quiet} additionally waits for completion and visibility~\cite{nvshmem_docs}.

This example illustrates that put/get is different from ordinary processor load/store. Put/get performs explicit memory-to-memory movement and may be asynchronous, even when it is invoked by device code. Its ordering and completion are defined by the communication interface rather than directly by the processor memory model. Operations issued through different communication queues or queue pairs may also require explicit synchronization.

AMD rocSHMEM exposes a related device-side model and includes tile-oriented put/get operations in addition to conventional one-sided transfers~\cite{rocSHMEM}. These implementations demonstrate that device initiation and operation semantics are separate choices: both host and device code may issue put/get, while device code may also initiate operations with different data granularities.

\smalltitle{GPU-initiated networking.}
GPU-Initiated Networking (GIN) extends device-side control from local scale-up links to network RDMA. The NCCL Device API allows communication to be initiated from GPU code and provides several execution paths. Load/Store Accessible (LSA) memory supports direct access over suitable NVLink or PCIe paths; Multimem exposes selected multicast and reduction capabilities; and GIN provides one-sided network transfers, signals, counters, and barriers
~\cite{NCCLDeviceAPI,hamidouche2025gpuinitiatednetworkingnccl}.

GIN supports both a kernel-initiated backend, in which the GPU interacts directly with a capable network endpoint, and a proxy backend, in which GPU code communicates with a host-side proxy through shared queues. The two paths can expose similar device interfaces but have different latency, CPU involvement, and hardware requirements. GPU-initiated communication should therefore not be interpreted as synonymous with a completely CPU-free implementation.

The main benefit of GIN is that communication can be issued and synchronized within a GPU kernel. This avoids repeated host scheduling, kernel launches, and CPU-GPU synchronization, and permits computation and communication to be fused more tightly. Such execution is particularly useful for MoE dispatch and combine, where routing decisions are produced dynamically by GPU computation and small communication operations may lie directly on the kernel critical path~\cite{hamidouche2025gpuinitiatednetworkingnccl}.

Importantly, GIN describes an initiation and execution model rather than a new data-movement semantic. The operations it realizes remain one-sided transfers, signals, counters, and synchronization. Similar device-side interfaces may select a local NVLink path, a PCIe path, or network RDMA according to endpoint reachability and topology.

\smalltitle{Tensor-oriented communication.}
Byte-oriented load/store and put/get do not directly express several properties of AI communication, including tensor layout, token routing, reduction, compression, and collective group membership. Recent interfaces therefore expose tensor- or tile-oriented operations. Examples include rocSHMEM tile put/get, NCCL device-side reduction and copy building blocks, the TMA/STMA-style data-movement mechanisms associated with Unified Bus systems, and the Tensor Push/Pull interface described by the ETH+-based UPN512 architecture ~\cite{rocSHMEM,NCCLDeviceAPI,UBSpec,ETHPlusUPN}.

A tensor-oriented operation may carry information about source and destination layouts, asynchronous or streaming execution, completion signals, reduction, compression, or in-network processing. These interfaces are particularly relevant to TP and EP because they can express application-level data movement more directly and reduce intermediate copies, kernel launches, and accelerator-memory traffic. They do not, however, eliminate the need to define ordering, completion, protection, and failure behavior.

\smalltitle{In-network collectives.} In-network collectives offload part of collective communication from hosts or accelerators to network components, the most common being switches and NICs. In the scale-up setting, I/O dies play a role analogous to that of NICs, but their relatively limited onboard resources can easily become a bottleneck for collective processing. Switch-executed collectives offer an alternative by placing this functionality within the fabric. This approach aligns with the broader trend toward lightweight accelerator endpoint stacks: simplifying endpoint protocols reduces communication overhead at the accelerators, offloading collective processing to switches further reduces the work performed by those endpoints. As a prime example, UALink adds optional in-network collective execution at two levels. Collective primitives replicate write and non-returning atomic requests, or reduce read responses, within the switch. Block collectives offload Broadcast, Reduce, and AllReduce: an accelerator submits a control block, after which the switch issues the required memory reads and writes, including accesses to contiguous or strided data layouts. A switch may implement neither collective level, primitives alone, or both primitives and block collectives. AllGather and ReduceScatter can be composed from repeated Broadcast and Reduce operations, respectively. Unlike ordinary unicast forwarding, collective execution requires switch-side group and transaction state, work queues, and multicast mechanisms.

Switch-executed collectives can reduce redundant data movement and endpoint processing, freeing accelerator resources for computation and shortening collective completion time. In-network reduction combines contributions before forwarding them, while multicast avoids repeated transmission of identical data over shared links. These broadcast- and reduction-oriented primitives directly benefit TP and DP; their benefits for EP and PP are more workload- and operation-dependent.

\smalltitle{Cross-system comparison.}
Table~\ref{tab:semantics} maps the eleven surveyed fabrics to their publicly documented transaction primitives and representative accelerator-visible realizations. The table distinguishes a complete platform interface from a fabric specification. A specification that defines read, write, or atomic transactions does not by itself establish that the accelerator ISA exposes corresponding remote load/store instructions. Three  observations follow.

The first observation is that operation semantics, initiation, and execution path are separate design dimensions. A put/get operation may be host- or device-initiated. A GPU-initiated operation may use direct load/store, a DMA engine, network RDMA, or a software proxy. Similarly, a fabric-level read/write transaction does not necessarily correspond to an accelerator instruction.

Second, systems in which one vendor controls the accelerator, memory subsystem, communication engines, software runtime, and interconnect can co-design these components to provide a shorter device-initiated communication path. This vertical integration can shorten the device-initiated path, but its complete ordering and visibility contract may not be publicly documented. Open fabric specifications typically define transaction behavior more explicitly while leaving the accelerator-facing API to implementations. Transport specifications such as TTPoE, and link/network specifications such as ESUN, do not prescribe the accelerator-visible operation interface, and we refrain from classifying them as providing memory semantics solely from their packet behavior. SUE-T instead defines a mapping for memory transactions, while leaving accelerator ISA and runtime integration implementation-specific.

Third, device-initiated execution offers a possible way to reduce the programming distinction between scale-up and scale-out communication. A kernel may issue a similar put/get, signal, or tensor operation while the runtime selects a local scale-up path or a network path. The underlying paths nevertheless retain different bandwidth, latency, ordering, and failure properties, which must remain visible to placement and scheduling mechanisms.

No single semantic model is optimal for all AI communication. Fine-grained synchronization and metadata access may benefit from load/store or atomics, whereas TP, EP, and DP commonly require asynchronous data movement and collective execution for large tensors. The relevant design question is therefore not simply whether a fabric supports "memory semantics", but whether its operation set, initiation path, and completion guarantees match the workload's communication critical path.

\subsection{Resource Sharing and Integration}
\label{subsec:resource-sharing}

As scale-up domains expand from a small number of tightly coupled accelerators to rack-scale systems, they increasingly become shared infrastructure rather than statically configured interconnects for a single workload. A scale-up fabric therefore addresses three related requirements: it divides physical resources among multiple tenants, prevents unauthorized or unintended cross-tenant access, and accommodates endpoints with different architectural and performance characteristics. Existing specifications provide several building blocks for these purposes, but their support varies considerably in granularity, enforcement scope, and degree of standardization.

\smalltitle{Virtualization and Access Isolation.} Existing scale-up specifications generally realize virtualization by constructing logical subsets of a physical fabric. UALink, for example, defines a \emph{Virtual Pod} as a tenant-owned subset of a physical pod. Accelerators assigned to a Virtual Pod are selected by the management plane, while switch routing tables restrict communication to the corresponding logical topology. An accelerator cannot belong to more than one Virtual Pod at a time, and accelerators outside a Virtual Pod are excluded from its routing configuration~\cite{UALink_spec}. While UALink permits SR\hyp{}IOV virtualization, each participating accelerator is limited to only one virtual function per domain and no fractional accelerator sharing is allowed. 


NVLink-based systems also support logical partitioning through the NVSwitch control plane and Fabric Manager. A physical NVLink domain can be divided into partitions such that GPUs and data paths in different partitions remain isolated \cite{nvlink_fabric_manager}. This is a platform-level capability rather than a fully open, implementation-independent protocol abstraction. Across these technologies, access isolation is therefore available, but it is commonly coarse-grained and dependent on topology or privileged management software.

\smalltitle{Trust boundary for collective offload.} Supporting in-network collectives can shift the trust boundary from ordinary accelerator endpoints to shared switches. For example, optional protection of collective traffic in UALink Common 2.0 brings participating switches into the tenant's trusted computing base after authentication and attestation. A switch decrypts incoming collective traffic, performs the operation, and encrypts outgoing collective traffic, while protected non-collective traffic remains encrypted through the switch. The Pod Controller and other infrastructure management software remain outside the tenant's trusted computing base~\cite{UALink_Common_2_0}. Thus, collective offload extends the security boundary beyond the accelerator endpoints, and in this case routing isolation alone is insufficient to describe the confidentiality guarantees.

\smalltitle{Performance Isolation and Quality of Service.}
Access isolation does not necessarily imply performance isolation. Even when two tenants cannot access each other's memory, their traffic may still contend for switch buffers, output ports, internal crossbars, or endpoint processing capacity. Existing specifications expose several mechanisms that can reduce such interference. UALink defines transaction-layer credit-based flow control and multiple virtual channels~\cite{UALink_spec}. Physical fabric partitioning can provide stronger isolation when links or switch resources are assigned exclusively to a partition.

These mechanisms nevertheless fall short of an end-to-end QoS abstraction. A virtual channel separates traffic classes and flow-control state, but does not by itself guarantee a tenant a minimum bandwidth or bounded tail latency. Similarly, endpoint-side resource partitioning, such as GPU multi-instance execution, can isolate compute and local-memory resources without controlling contention inside the scale-up fabric. Current specifications generally leave admission control, bandwidth allocation, arbitration policies, and the mapping from workload objectives to fabric resources to implementations or system software. Consequently, portable guarantees for bandwidth, latency, and collective-operation completion time remain limited.

\smalltitle{Heterogeneous Endpoint Integration.}
The ability to connect heterogeneous endpoints is another increasingly important property. UALink instead defines a common scale-up interface for accelerators and is intended to permit accelerators and switches from different vendors to participate in the same pod~\cite{UALink_spec}. NVLink has traditionally been more closely coupled to NVIDIA GPUs and system platforms, although newer platform initiatives extend connectivity toward custom CPUs and accelerators~\cite{nvlink_fabric_manager}.

Protocol compatibility, however, is only one level of heterogeneous integration. Endpoints may differ in memory consistency models, atomic operations, address-translation mechanisms, supported ordering rules, collective offload capabilities, failure behavior, and achievable injection rates. A common physical and transaction interface does not automatically reconcile these differences. In particular, the overall behavior of a heterogeneous scale-up domain may be constrained by the least capable endpoint, or may require software to discover and manage device-specific capabilities. Existing specifications thus provide
increasingly broad endpoint connectivity, but offer less support for portable semantic and performance behavior across heterogeneous devices.

\smalltitle{Summary.} Overall, current scale-up specifications provide meaningful mechanisms for logical partitioning, access control, traffic separation, and multi-vendor device connectivity. Their strengths are primarily in constructing isolated logical fabrics and defining interoperable transaction paths. They provide substantially weaker guarantees for fine-grained resource sharing, end-to-end performance isolation, and semantic portability across heterogeneous endpoints. This distinction
is important: the remaining challenge is not the complete absence of virtualization or heterogeneity support, but the lack of a unified abstraction that connects tenant-level policies to routing, protection, bandwidth allocation, and endpoint capabilities.
\section{Challenges and Open Problems}
\label{sec:challenges}
\begin{figure*}
    \centering
    \includegraphics[width=0.7\linewidth]{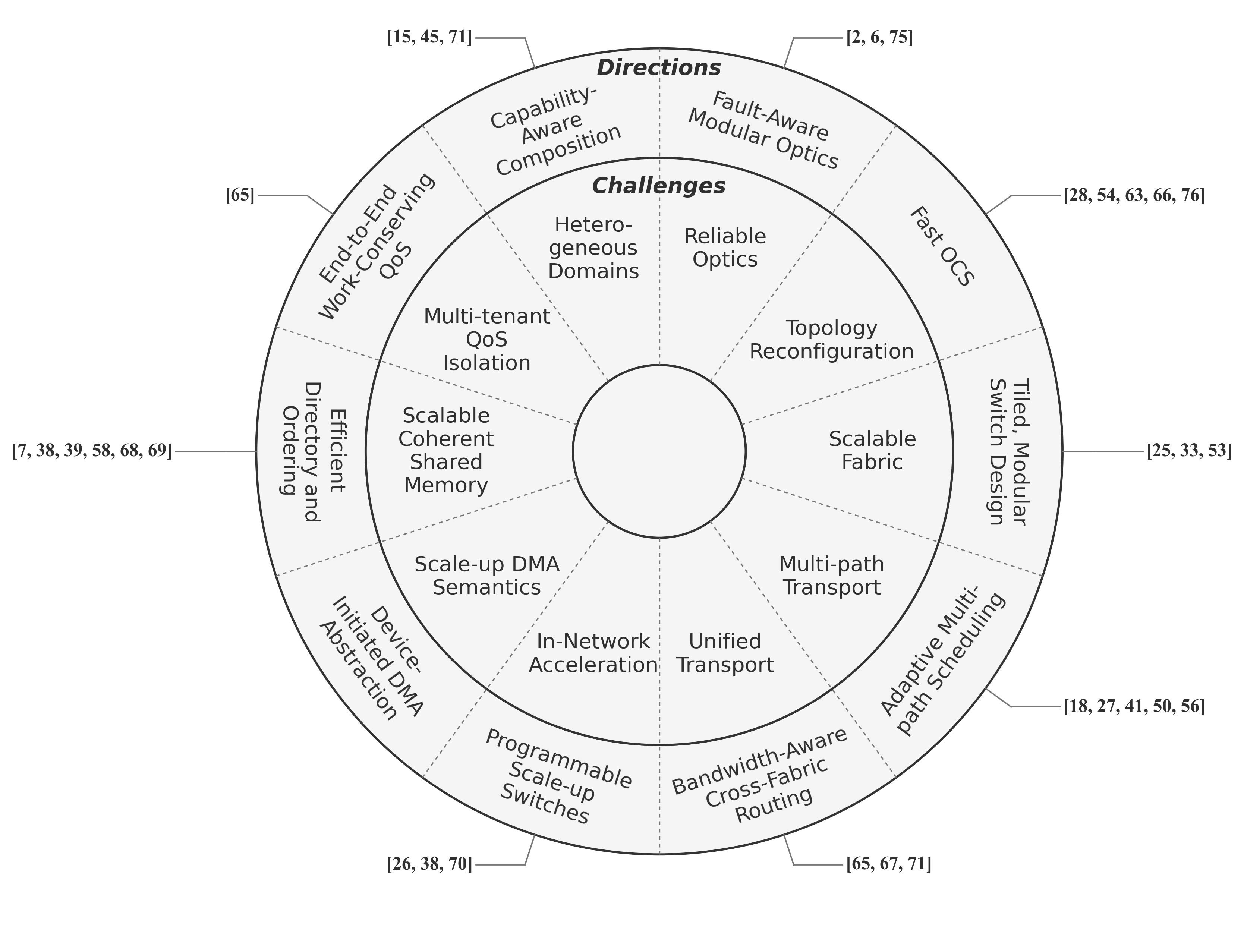}
    \caption{Challenges and research directions for next-generation scale-up fabrics}
    \label{fig:wheel}
\end{figure*}

To ground our analysis in practical experience, we consulted multiple industry experts on the unresolved issues surrounding scale-up networking. Drawing on these discussions, this section identifies the key challenges and open problems that have yet to reach broad consensus in industry or academia. Those span four dimensions: scaling the underlying fabric, efficiently utilizing its communication capacity, exposing tightly coupled system semantics, and sharing and composing the resulting scale-up domain. The challenges and their corresponding future research directions are summarized in Figure \ref{fig:wheel}.

\subsection{Physical Interconnect and Switching Fabric}
\smalltitle{Challenge 1: Reliability and Cost of Optical Interconnects.}
As scale-up domains grow, packaging, power, and thermal constraints make it increasingly difficult to keep all accelerators within a single rack. At the same time, the reach of high-speed passive copper links decreases as signaling rates rise. For example, at 200~Gb/s per lane, passive copper cannot span typical cross-rack distances without electrical retimers, which introduce additional power and latency overheads \cite{BroadcomCPOBiDi}. Optical interconnects therefore become a natural choice for extending scale-up fabrics across racks.

Compared with short-reach passive copper links, conventional active optical cables used in today's data centers can consume more power and have been reported to exhibit failure rates up to two orders of magnitude higher~ \cite{Mosaic}.
This transition, however, introduces additional reliability and cost challenges.
In a densely connected scale-up fabric, failures of these components can cause link flapping, imbalanced link utilization, packet loss, etc. \cite{Zhuo2017}. These effects are particularly damaging because scale-up workloads rely on high bandwidth and predictable tail latency. Moreover, optical engines, transceivers, fiber infrastructure, and maintenance increase deployment and operational costs, making extensive use of optics economically challenging.

The open problem is therefore how to provide cross-rack optical reach without sacrificing the reliability, latency, and cost efficiency expected from a scale-up fabric. Promising directions include lower-power and more reliable optical devices, modular and serviceable transceivers, proactive failure monitoring, and lightweight fault-aware redundancy. Such mechanisms must react quickly to component degradation while adding little control-plane complexity or data-path overhead.


\begin{figure}[t]
    \centering

    \subfloat[
        Full mesh, All-to-All \\
        ($28/28$ links, max hop = 1).
        \label{fig:fullmesh-alltoall}
    ]{
        \includegraphics[width=0.4\linewidth]
        {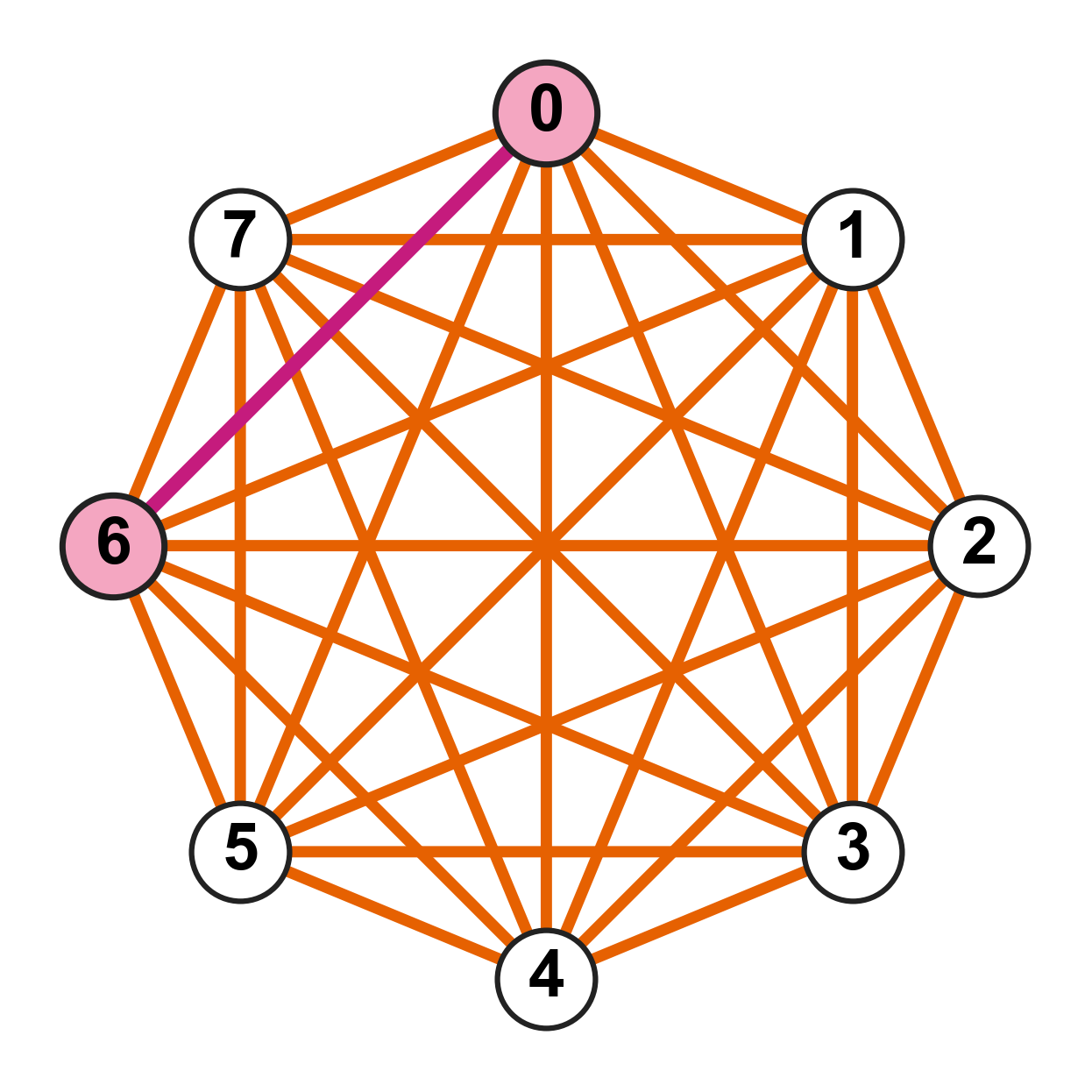}
    }
    \hfill
    \subfloat[
        2D torus, All-to-All \\
        ($16/16$ links, max hop = 3).
        \label{fig:torus-alltoall}
    ]{
        \includegraphics[width=0.4\linewidth]{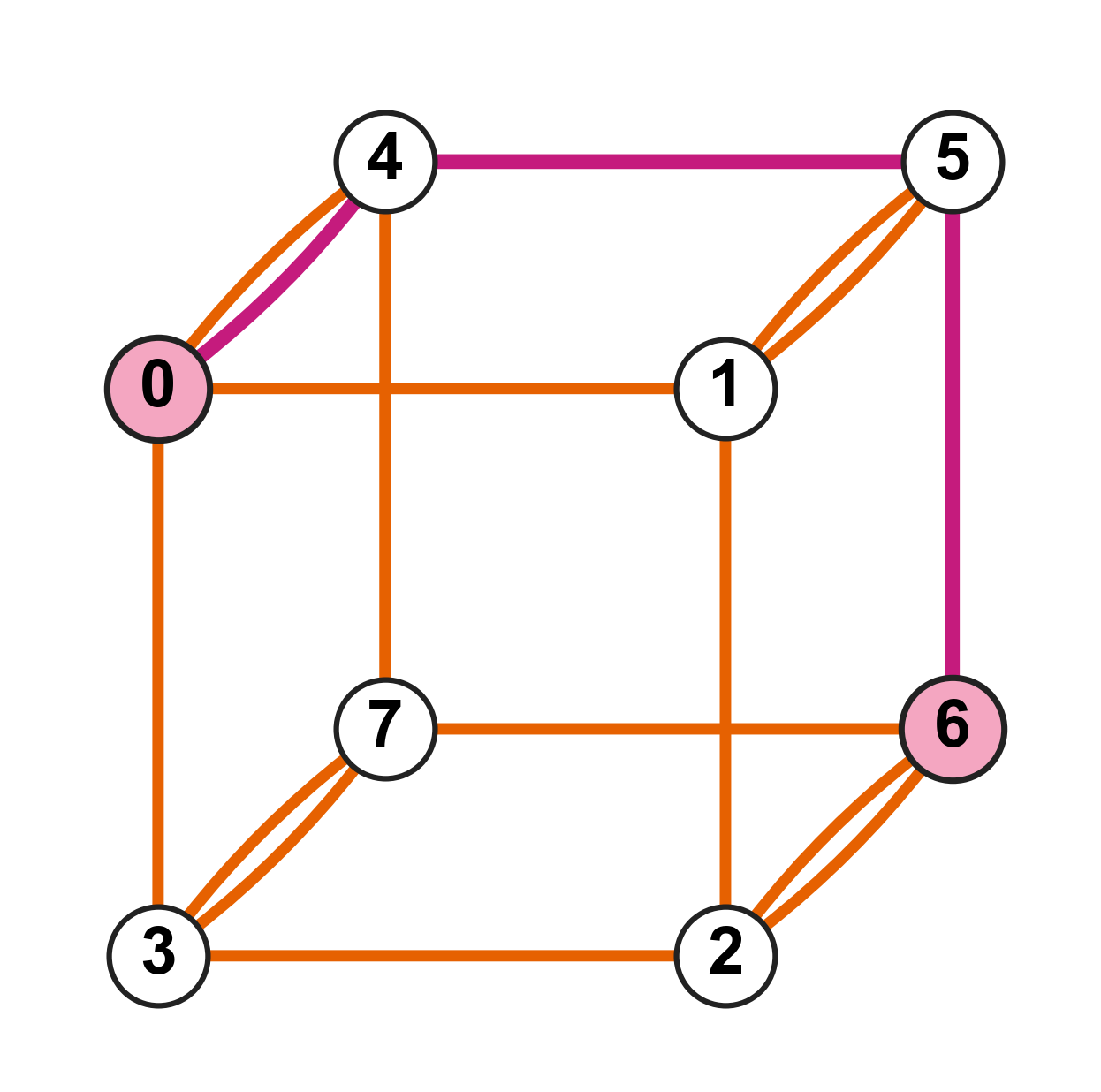}
    }
    \hfill
    
    \subfloat[
        Full mesh, ring AllReduce
        \\ ($8/28$ links).
        \label{fig:fullmesh-ring}
    ]{
        \includegraphics[width=0.4\linewidth]
        {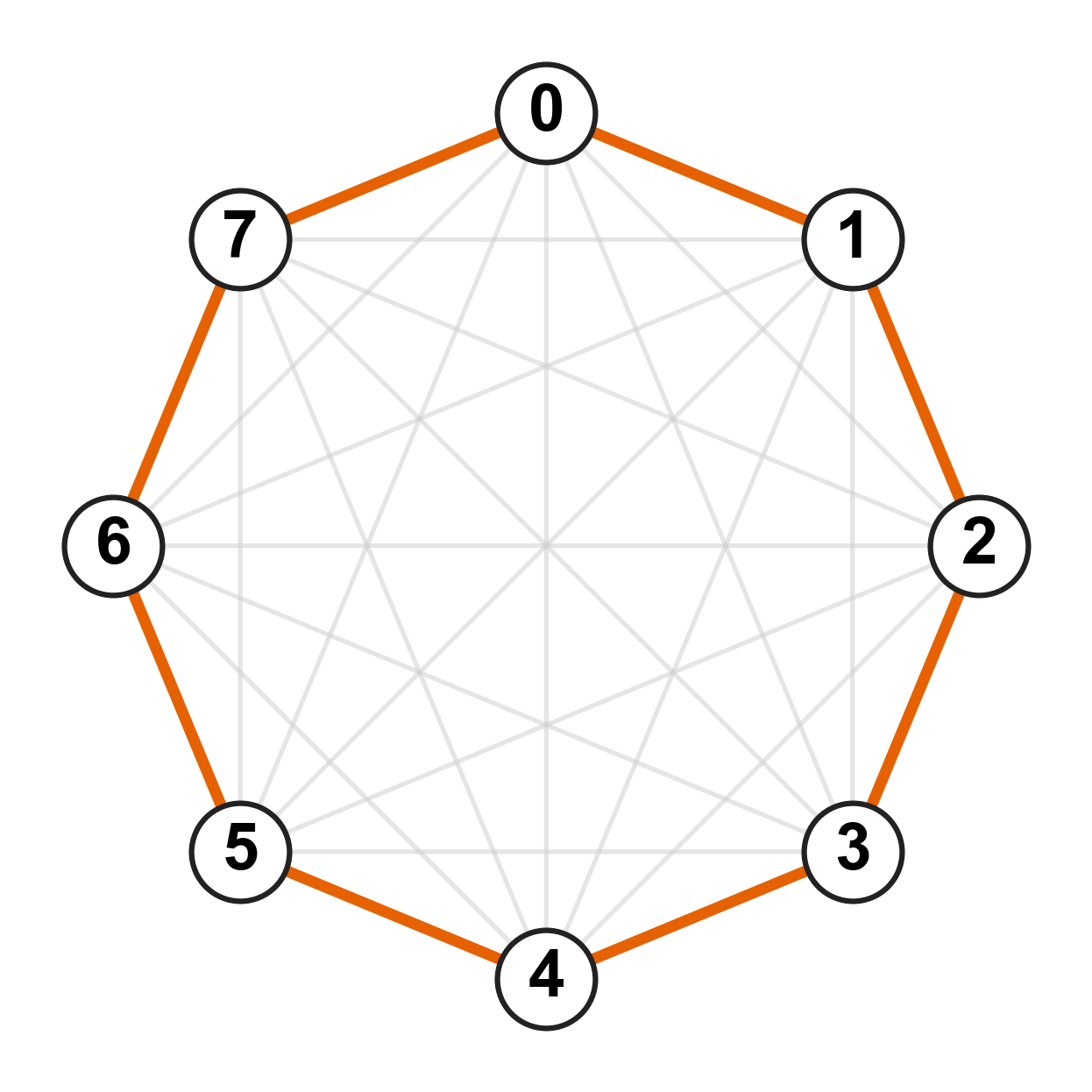}
    }
    \hfill
    \subfloat[
        2D torus, AllReduce \\
        ($16/16$ links).
        \label{fig:torus-allreduce}
    ]{
        \includegraphics[width=0.4\linewidth]
        {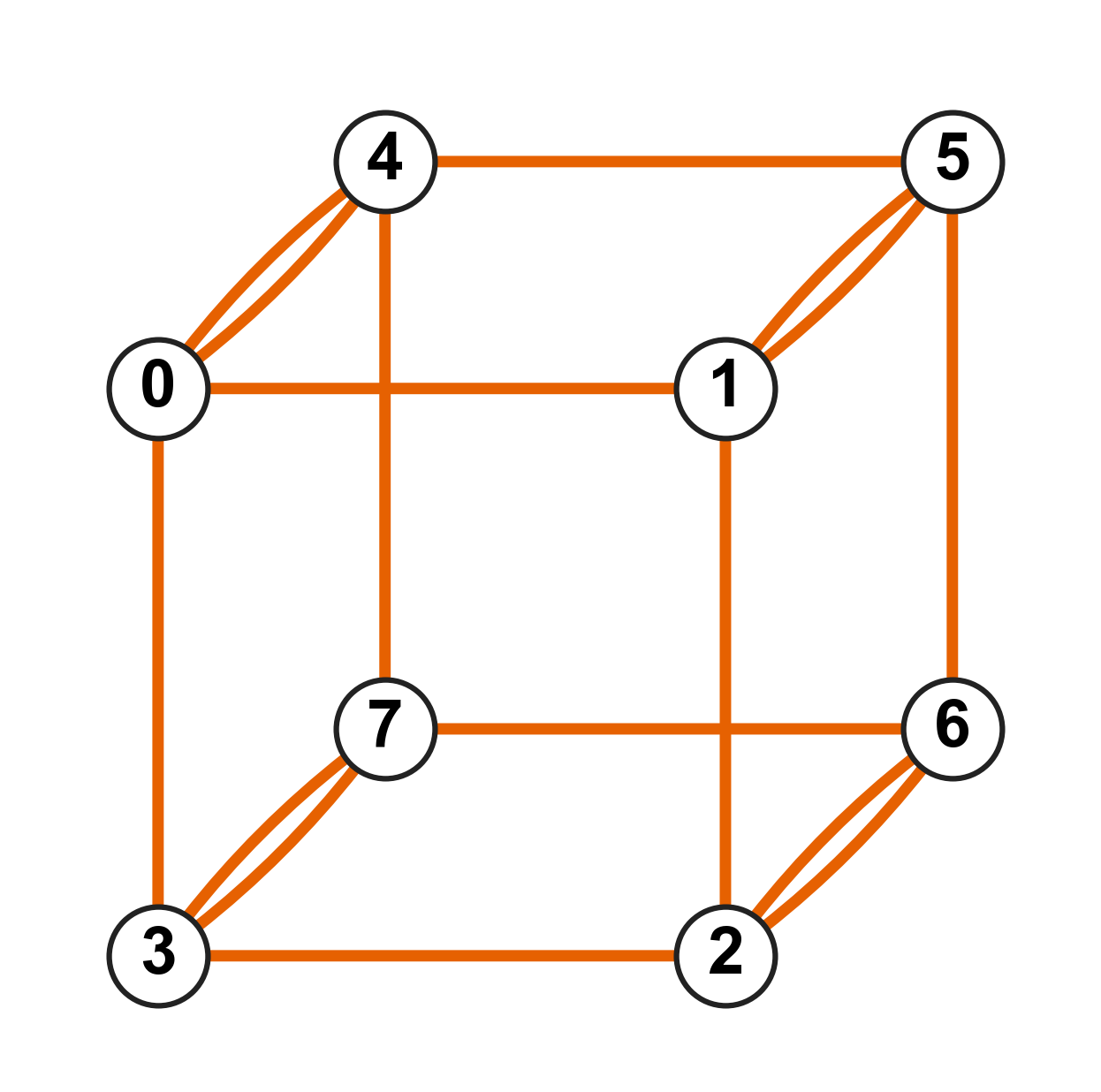}
    }

    \vspace{0.55em}

\begin{minipage}{\linewidth}
\raggedright

{\small
\begin{tabular}{c}
    \begin{tikzpicture}[baseline=-0.5ex]
        \draw[usedlink, line width=1.5pt]
            (0,0) -- (0.65,0);
        \node[anchor=west] at (0.78,0) {Used link};

        \draw[unusedlink, line width=1.1pt]
            (2.75,0) -- (3.40,0);
        \node[anchor=west] at (3.53,0) {Unused link};
    \end{tikzpicture}
    \\[0.28em]
    \begin{tikzpicture}[baseline=-0.5ex]
    \filldraw[
        fill=highlightnode,
        draw=black,
        line width=0.5pt
    ] (0.12,0) circle (0.115);
    \node[anchor=west] at (0.35,0) {Highlighted Accelerators 0 and 6};

    \draw[highlightlink, line width=1.8pt]
        (5,0) -- (5.65,0);
    \node[anchor=west] at (5.7,0)
        { $0\!\leftrightarrow\!6$ paths};
\end{tikzpicture}

\end{tabular}
}
\end{minipage}

   \caption{Comparison of full-mesh and 2D torus topologies for All-to-All and ring AllReduce Communications.}
    \label{fig:topo_comp}
\end{figure}

\smalltitle{Challenge 2: Topology Reconfiguration through Optical Switching.}
The communication pattern varies across workloads due to the use of different collective operation and parallelization strategies. As a result, mismatch between the fabric topology and workload is possible and can cause link under-utilization or increased latency. A typical example is shown in Figure \ref{fig:topo_comp}: the links in a full-mesh topology match All-to-All communication but suffer substantial idleness under ring AllReduce; by contrast, the 2D torus incurs a maximum hop count of three for All-to-All communication \cite{Daniel2024}.


Topology reconfiguration allows the fabric to reshape its connectivity according to the current communication demand. By establishing more direct or higher-capacity connections between frequently communicating accelerators, it can reduce hop count and contention, improve link utilization, and consequently lower communication latency. ZCube and TopoOpt demonstrate the performance benefits of constructing topologies around workload communication patterns \cite{ZCube,TopoOpt}, while MixNet applies runtime reconfiguration to dynamic MoE traffic \cite{MixNet}.

Reconfiguration can also improve cost efficiency. Instead of provisioning a fully connected or high-bandwidth topology for every possible traffic pattern, the same pool of physical links can be reassigned to the connections that currently require them. Similarly, after component failures, healthy accelerators and links can be recomposed into usable domains, allowing backup capacity to be shared rather than statically reserved for individual failure domains. TPUv4 demonstrates the value of such reconfigurable connectivity for retaining healthy resources under failures \cite{TPUv4}. 

The open problem is how to realize these performance and cost benefits with sufficiently low reconfiguration latency, control complexity, and additional hardware overhead.

\smalltitle{Challenge 3: Low-Latency Switching at Scale.}
As a scale-up domain grows, the limited link radix of each accelerator makes switched and potentially multi-stage fabrics necessary. Contemporary NVLink systems, for example, use multiple NVLink switches to provide high-bandwidth connectivity across the accelerator domain~\cite{Nvidia_NVSwitch}. However, increasing the number of switches also increases the number of forwarding, arbitration, and queuing stages on the communication path. This additional latency is particularly significant for fine-grained memory operations and latency-sensitive collectives.

Conventional packet switches commonly perform several stages of parsing, lookup, buffering, scheduling, and forwarding. Such a pipeline may be too long for a tightly coupled scale-up fabric, especially when multiple switch hops must be traversed. Deep buffers are also undesirable because they consume substantial on-chip memory and can increase both buffer-access and queuing latency. Cut-through or flit-level forwarding provides a promising alternative: a switch can begin forwarding a message after receiving sufficient routing information, thereby overlapping reception and transmission rather than waiting for the entire packet~\cite{virtual_cut_through}. Nevertheless, shallow-buffer cut-through fabrics remain vulnerable to output contention, head-of-line blocking, and transient incast, while backpressure can propagate across multiple switches.

The open problem is therefore how to preserve low and predictable end-to-end latency as the scale-up switch fabric grows in radix and hop count. This requires network-level mechanisms beyond simply building a larger crossbar, including lightweight fast paths, lookahead or speculative arbitration, flit-level scheduling, switch bypass, adaptive routing, and rapid local congestion feedback. These mechanisms must sustain bursty collective and memory traffic with small buffers while preserving lossless delivery, ordering, and deadlock freedom. The central challenge is to determine how much switch processing can be removed from the critical path, and how routing, arbitration, and flow control should be co-designed so that a scalable or Ethernet-compatible fabric can approach the latency of a small dedicated scale-up switch.

\subsection{Efficient Communication}

\begin{figure}[t]
    \centering
    \includegraphics[width=\linewidth]{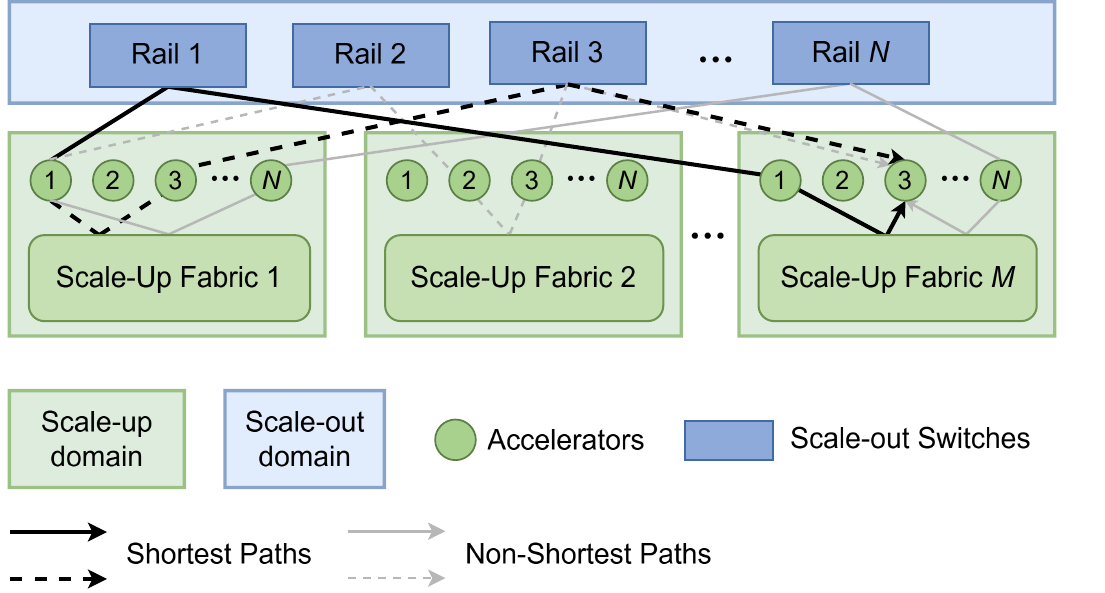}
    \caption{Sample inter-scale-up-domain paths}
    \label{fig:path}
\end{figure}

\smalltitle{Challenge 4: Multi-Path Transport for Scale-Up Fabrics.} 

Scale-up endpoints commonly expose multiple high-bandwidth ports, while switched topologies may provide multiple paths between accelerator pairs. Concurrent use of these ports and paths can aggregate bandwidth, improve resilience, and reduce dependence on an individual link or switch. However, path diversity alone does not guarantee high goodput. Traffic imbalance, shared bottlenecks, heterogeneous path delays, receiver-side resource constraints, and interactions with congestion control can leave some paths underutilized while overloading others. Similar challenges have been observed in data-center and scale-out networks~\cite{Li2016}, including systems that aggregate multiple NICs for GPU communication~\cite{10.5555/3767901.3767907}.

Whether multi-path transport can freely reorder traffic depends on the ordering contract exposed by the fabric. Some scale-up interfaces provide relaxed or non-strict ordering modes for independent transactions, while still requiring ordering within a queue, stream, or explicitly synchronized scope. Other operations, including atomics, synchronization messages, and dependent memory accesses, may require stronger ordering. Multi-path transport can exploit the reordering permitted by a particular operation mode, but it cannot assume that memory-semantic traffic is inherently weakly ordered.

Preserving these contracts across paths with different delays may require sequence tracking, receiver-side reordering, or coordination at fences and other synchronization boundaries. Alternatively, blocking later transactions until all earlier operations complete can simplify correctness but may drain the transmission pipeline and reduce aggregate bandwidth. Recent scale-up proposals expose this trade-off between reordering state and synchronization-induced blocking~\cite{FARE}. Existing multi-path transports such as Falcon and MRC further demonstrate that path selection must be coordinated with congestion control, reliable delivery, and ordering semantics~\cite{Falcon,MRC}.

The open problem is how to utilize the aggregate bandwidth of multiple scale-up ports and paths while preserving the ordering and completion guarantees required by each operation class. A practical design must limit reordering state and synchronization-induced blocking, remain responsive to congestion and failures, and expose sufficiently clear semantics for endpoint hardware and software.

\smalltitle{Challenge 5: Sustaining Fine-Grained Accelerator-Initiated Data Movement.}
The bandwidth provided by a scale-up fabric can be utilized only if accelerator endpoints can generate, track, and complete data-movement operations at a sufficient rate. This becomes challenging as per-accelerator fabric bandwidth approaches the TB/s range while many transfers remain fine-grained. In this regime, the endpoint request path, rather than the physical link, may become the performance bottleneck.

Processor-issued \texttt{load}/\texttt{store} operations provide tight integration with registers and the processor memory model. However, a remote load places the fabric latency directly on the dependency path of the issuing thread, and the number of concurrent operations is constrained by processor and memory-system resources. At the other extreme, DMA-like engines move data asynchronously between memories and amortize their setup cost over larger blocks. Nevertheless, descriptor generation, queue management, address translation, request segmentation, and completion tracking can limit their operation rate, particularly for small transfers. Accelerator-local mechanisms such as NVIDIA's Tensor Memory Accelerator (TMA) demonstrate how address generation and bulk tensor movement can be offloaded from compute cores, but TMA itself is not a remote-network communication semantic~\cite{NVIDIATMA}.

One-sided \texttt{put}/\texttt{get} operations represent a possible intermediate abstraction. They move data between local and remote accelerator memories, can be issued asynchronously by an accelerator, and do not require the remote processor or, in a hardware-native implementation, the host CPU, to participate in the per-operation data path. Although \texttt{put}/\texttt{get} is not a new programming model, providing it as a lightweight, accelerator-native scale-up primitive raises new design questions. These include the maximum operation size and alignment, segmentation into fabric transactions, addressing and protection, ordering and visibility, completion notification, backpressure, and the number of outstanding operations.

The open problem is how endpoints should combine synchronous \texttt{load}/\texttt{store}, fine-grained asynchronous \texttt{put}/\texttt{get}, and bulk data-movement engines without allowing request-generation or completion-processing overhead to become the bottleneck.

\smalltitle{Challenge 6:  In-Network Acceleration for Scale-Up Collectives. } 
In-network computation has been explored in scale-out networks to reduce collective traffic and endpoint processing~\cite{switchml,sharp}. However, its general deployment remains a challenge. Commodity network switches provide limited computation and on-chip memory, while collective offloading must additionally handle congestion and coordination across multi-level network topologies. Consequently, existing solutions often require specialized switches, transport modifications, or tightly controlled deployments.

Scale-up domains provide a more supportable environment for in-network collective acceleration. They typically consist of a bounded set of accelerators connected through a tightly coupled, low-latency fabric with a relatively regular topology. The switches and endpoints are commonly co-designed and managed as a single system, while tensor-parallel workloads repeatedly invoke predictable collectives among stable groups of accelerators. These properties make it more practical to integrate collective engines into scale-up switches, reduce redundant data movement, and coordinate computation with memory-semantic communication~\cite{klenk2020inc,nvls}. 

Recent work such as DySHARP further shows that scale-up In-Network Computing (INC) can be extended beyond static collectives to dynamic and irregular MoE communication~\cite{dysharp}. ETH+ moves in the same direction, standardising in-fabric collective semantics (IFEC) that offload Reduce, Broadcast, ReduceScatter, AllGather, and AllToAll to switches or network interfaces~\cite{IFEC}. Nevertheless, several open issues remain, including which collective operations should be offloaded, how limited switch computation and buffering should be shared, and how to support different data types, irregular communication patterns, synchronization, and failures without compromising the primary forwarding function of the fabric.



\subsection{System Semantics}


\smalltitle{Challenge 7: Accelerator-Initiated Communication across Scale-Up and Scale-Out Fabrics.}
\label{sec:challenge-gpu-initiated} Modern AI kernels increasingly interleave computation with communication. However, many communication stacks remain host mediated: an accelerator kernel produces data, returns control to the host, and waits for the runtime to submit the corresponding communication operation. The resulting device-to-host notification, synchronization, command submission, and kernel relaunch introduce gaps between computation and communication. These overheads are particularly undesirable for fine-grained transfers and latency-sensitive inference, where they may lie directly on the execution critical path.

The problem becomes more pronounced when an AI workload spans both scale-up and scale-out fabrics. A representative example is hierarchical gradient synchronization. Accelerators first reduce gradients among peers within a scale-up domain, exploiting its high bandwidth and low latency. The partially reduced results are then exchanged among data-parallel replicas in different domains through a scale-out network, followed by a final intra-domain distribution. Therefore, a single logical collective traverses two fabrics with different bandwidth, latency, addressing, transport, and reliability characteristics.

Ideally, the accelerator-side communication kernel should be able to initiate and compose these operations without returning control to the host at each fabric boundary. From the kernel's perspective, an intra-domain transfer and an inter-domain transfer may be expressed
through similar operations, such as put/get, signal, or collective operations. The underlying hardware and runtime can then select the
appropriate scale-up or scale-out path according to the destination. This does not require the two fabrics to use identical physical or transport protocols. Instead, it requires a common accelerator-visible communication abstraction with sufficiently consistent addressing, ordering, completion, and synchronization rules.

Such an abstraction can also improve communication efficiency. A persistent or fused kernel may produce a data block, initiate its
transfer through the appropriate fabric, continue executing independent work, and subsequently observe device-visible completion state. This execution model can reduce host round trips and expose additional opportunities to overlap computation, intra-domain communication, and inter-domain communication. It is especially relevant to kernel and communication-library developers, because the communication operations must be composable with kernel execution rather than available only as host-side library calls.

Several questions remain open. First, the interface should define consistent ordering and completion semantics when operations traverse
different communication engines and fabrics. Second, the accelerator should make communication progress without frequent host intervention, while avoiding excessive consumption of execution and memory-system resources. Finally, the system should efficiently support both fine-grained control operations and bandwidth-oriented bulk transfers. Addressing these issues is essential to making accelerator-initiated communication both efficient within a scale-up domain and composable across scale-up and scale-out networks.

\smalltitle{Challenge 8: Coherent Shared Memory over Scale-Up Fabrics.} Cache coherence is an important enabler for shared memory across a scale-up domain. Shared memory allows AI accelerators to access common model and inference states without maintaining a separate copy at each endpoint or repeatedly moving data between devices. Existing systems have explored global or fabric-attached memory for fine-grained GPU communication, large embedding tables, and shared KV-cache management~\cite{nvshmem_docs,carina}. When multiple accelerators cache and update such shared data concurrently, coherence prevents them from observing stale values and enables transparent
load/store access without explicit flush, invalidation, or data-copy operations. As shared-memory abstractions become more
transparent and allow concurrent cached access from heterogeneous accelerators, maintaining coherence becomes an increasingly important requirement of the scale-up fabric.

Supporting coherence across a scale-up domain introduces several network challenges. Broadcast-based snooping generates traffic that grows with the number of endpoints, whereas directory-based protocols require substantial metadata to track the sharers and owner of each cache line~\cite{taglessdirectory}. A coherence transaction may also involve multiple dependent messages, including requests, snoops, invalidations, acknowledgements, and data responses. The network must preserve their required ordering and provide deadlock-free progress, often through separate virtual channels or virtual networks~\cite{coherencevn}. Concurrent accesses to the same cache line can cause multiple coherence transactions to overlap. Since each transaction consists of dependent requests, snoops, data responses, and acknowledgements, message reordering and protocol-level retries create numerous transient states that must be handled correctly to prevent stale data or conflicting ownership~\cite{formalisingcxl}.

The open problem is how to provide low-latency and bandwidth-efficient coherence across hundreds of accelerators without excessive directory storage, coherence traffic, or network-level ordering constraints.

\subsection{Sharing and Composing the Scale-Up Domain}



\smalltitle{Challenge 9: Multi-Tenant QoS Isolation in Scale-Up Fabrics.}
Multi-tenancy improves the utilization and cost efficiency of accelerator clusters by allowing several training or inference jobs to share a scale-up domain. However, these jobs may have substantially different communication patterns and performance objectives. For example, latency-sensitive inference can coexist with bandwidth-intensive training, while collective operations from one tenant may generate synchronized traffic bursts across many accelerators. Without QoS isolation, an aggressive tenant can occupy shared links and switch buffers, causing unpredictable collective completion times, tail-latency violations, or even starvation for other tenants. Therefore, the scale-up fabric must provide predictable performance in addition to basic connectivity and bandwidth~\cite{mccs}.

Providing such isolation is challenging because scale-up fabrics operate at extremely high link rates and low latency, leaving little time and on-chip state for complex per-tenant scheduling. Moreover, a tenant's communication demand is distributed across multiple endpoints, paths, and switches. Local scheduling at a single output port cannot guarantee end-to-end bandwidth or latency when collective flows contend at several locations. The synchronized and many-to-many nature of collective communication further makes traffic bursty, while receiver contention and backpressure can propagate interference across the domain. Finally, strict priority can protect latency-critical traffic but may starve lower-priority tenants, whereas static bandwidth reservations can waste capacity when reserved resources are idle.

The open problem is how to provide end-to-end, work-conserving QoS isolation across a scale-up domain with sufficiently low hardware and latency overhead. A practical design must coordinate admission control, traffic classification, scheduling, and congestion control across endpoints and switches, while providing predictable latency or bandwidth guarantees without sacrificing fabric utilization or collective progress.


\smalltitle{Challenge 10: Composable Heterogeneous Scale-Up Domains.}
Production scale-up accelerator domains are typically composed of homogeneous accelerator endpoints. Such a design simplifies system integration and collective communication, but fixes the ratio between compute capability, memory capacity, and memory bandwidth at the domain level. This fixed resource composition may not match the diverse requirements of AI workloads. For example, the prefill phase of LLM inference is relatively compute-intensive, whereas decoding is more sensitive to memory capacity and bandwidth. Prior work has shown that placing these phases on differently provisioned resources can improve throughput and cost efficiency~\cite{splitwise,DistServe}. A heterogeneous scale-up domain could similarly combine compute-rich, memory-rich, and potentially memory-only endpoints, allowing workloads to be placed on resources that better match their characteristics.

Realizing this design, however, introduces several challenges. Accelerators from different vendors or generations may expose incompatible physical interfaces, addressing models, memory semantics, atomic operations, and collective-communication libraries. Even when interoperability is provided by a common fabric, heterogeneous endpoints can have substantially different compute speeds, memory bandwidths, and traffic-injection rates. Conventional collective algorithms and tensor-partitioning strategies often assume symmetric participants; under heterogeneity, faster endpoints may wait for slower ones, while equal traffic allocation can overload low-bandwidth links or leave high-bandwidth links underutilized. Recent work on heterogeneous collective communication demonstrates the importance of capability-aware communication and topology-aware scheduling~\cite{heteccl,forestcoll}.

The open problem is how to build an interoperable and composable scale-up domain that can efficiently coordinate heterogeneous accelerators and memory resources. Practical solutions have to discover endpoint capabilities, expose common communication semantics, and jointly determine workload placement, data partitioning, and collective schedules according to asymmetric compute and communication capacities.

\section{Conclusion}


Scale-up networking is becoming a distinct system-design domain rather than merely a higher-bandwidth form of scale-out networking. Its performance depends on the joint design of the interconnect, accelerator architecture, memory system, collective library, and runtime. Accordingly, topology or peak bandwidth alone cannot determine the capability of a scale-up domain.

Our comparison of eleven representative solutions yields three main observations. First, achievable domain scale is jointly constrained by endpoint connectivity, switch radix, link reach, oversubscription, routing, and transport mechanisms. Second, the scale-up solutions cover different functional layers: some specify complete system architectures, while others define only links, transports, transaction interfaces, or switching functions. Third, vertical co-design can reduce communication overhead and support accelerator-specific operations, but may limit interoperability, modularity, and implementation transparency.

Future scale-up systems must address more than link bandwidth. Key open problems include predictable tail latency, congestion control, accelerator-initiated communication with explicit completion and visibility semantics, performance isolation, heterogeneous memory access, and composition across multiple scale-up domains. Although the analysis is limited by uneven and rapidly evolving public documentation, the proposed workload-driven and layer-aware framework provides a consistent basis for evaluating scale-up technologies and identifying their remaining architectural gaps.




\bibliographystyle{ACM-Reference-Format}
\bibliography{main}

\end{document}